\documentclass[sigconf]{acmart}
\AtBeginDocument{%
  \providecommand\BibTeX{{%
    \normalfont B\kern-0.5em{\scshape i\kern-0.25em b}\kern-0.8em\TeX}}}
\usepackage{amsmath,amsfonts}
\usepackage{algorithmic}
\usepackage{graphicx}
\usepackage[caption=false]{subfig}
\usepackage{float}
\usepackage{multirow}
\usepackage{array}
\graphicspath{{figs/}}
\usepackage{url}
\usepackage{caption}
\copyrightyear{2022} 
\acmYear{2022} 
\setcopyright{rightsretained} 
\acmConference[SNTA'22]{Proceedings of the Fifth International Workshop on Systems and Network Telemetry and Analytics}{June 30, 2022}{Minneapolis, MN, USA}
\acmBooktitle{Proceedings of the Fifth International Workshop on Systems and Network Telemetry and Analytics (SNTA'22), June 30, 2022, Minneapolis, MN, USA}
\acmDOI{10.1145/3526064.3534110}
\acmISBN{978-1-4503-9315-7/22/06}

\begin{document}
\fancyhead{}
\title{Access Trends of In-network Cache for Scientific Data}

\author{Ruize Han}
\affiliation{%
  \institution{University of California, Berkeley}
  \city{Berkeley}
  \state{CA}
  \country{USA}
}
\email{hrz98@berkeley.edu}

\author{Alex Sim}
\author{Kesheng Wu}
\affiliation{%
  \institution{Lawrence Berkeley Nat'l Laboratory}
  \city{Berkeley}
  \state{CA}
  \country{USA}
}
\email{{asim,kwu}@lbl.gov}

\author{Inder Monga}
\author{Chin Guok}
\affiliation{%
  \institution{Energy Sciences Network}
  \city{Berkeley}
  \state{CA}
  \country{USA}
}
\email{{imonga,chin}@es.net}

\author{Frank~W\"{u}rthwein}
\author{Diego Davila}
\affiliation{%
  \institution{University of California, San Diego}
  \city{La Jolla}
  \state{CA}
  \country{USA}
}
\email{{fkw,didavila}@ucsd.edu}

\author{Justas Balcas}
\author{Harvey Newman}
\affiliation{%
  \institution{California Institute of Technology}
  \city{Pasadena}
  \state{CA}
  \country{USA}
}
\email{{jbalcas,newman}@hep.caltech.edu}

\renewcommand{\shortauthors}{Han, et al.}

\begin{abstract}
Scientific collaborations are increasingly relying on large volumes of data for their work and many of them employ tiered systems to replicate the data to their worldwide user communities.
%Consequently the network bandwidth requirement increases to replicate the data as well as for users to access the data.
Each user in the community often selects a different subset of data for their analysis tasks; 
however, members of a research group often are working on related research topics that require similar data objects.
Thus, there is a significant amount of data sharing possible.
In this work, we study the access traces of a federated storage cache known as the Southern California Petabyte Scale Cache.
By studying the access patterns and potential for network traffic reduction by this caching system, we aim to explore the predictability of the cache uses and the potential for a more general in-network data caching.
Our study shows that this distributed storage cache is able to reduce the network traffic volume by a factor of 2.35 during a part of the study period.
%\fix{This description of network traffic reduction is a little unfamiliar to me, suggest to use a more common description.  For example, if the original traffic was 100\%, we might say the traffic was reduce by 60\% (meaning the new reduced traffic is 40\% of the original).  Another possible description might be the new traffic volume is 1/X of the original.  My suspicion is that you want to say the new traffic is 1/(1+1.30), but this might be too convoluted to express?? It's described in section 3.}
We further show that machine learning models could predict cache utilization with an accuracy of 0.88.
This demonstrates that such cache usage is predictable, which could be useful for managing complex networking resources such as in-network caching.
%where 80\% and 89\% of the true values fall into the 2 STDs and 3 STDs of the predicted values respectively. 
%These results would support the sustainability of the data access for the in-network caching.
%\fix{These last two sentences could be more specific.  For example. we have a develop a LSTM model that could predict the cache hit rate to 90\% accuracy. [AS] Jack, do we have the performance number besides RMSE number?}
%\fix{Will need to read through the read of paper to say more about the connection between the above predictions and in-network cache.  My thinking is that in-network cache is used by the internet backbone service providers to reduce the traffic on Wide-Area Network.  It is offered as a general service rather than a data lake operated by a specific user community.  The existing evidence shows evidence of reduce traffic on WAN.  However the predictions come into this? Prediction is for the network bandwidth planning. If you can predict how much bandwidth a regional cache would use in about a year, it'll be heplful planning in advance.}

%\fix{need to say something about prediction and in-network cache}.
\end{abstract}




\maketitle

\section{Introduction}
\label{sec:intro}
% 1: distributing data to a large user community
The increasing volume of data from scientific experiments and simulations requires a vast amount of resources to store and distribute to geographically distributed users.
Many collaborations such as the Large Hadron Collider (LHC) utilize tiered systems to replicate the data in a few places, and the users could access their nearby storage sites.
However, with the increasing cost of managing storage resources and the limited number of replicas, the large number of user accesses still create considerable demand on the wide-area network that increases the cost of data analyses, and could cause large-scale network traffic congestion~\cite{datalakes, esnetHepReq}.

% 2. caching reduces access latency and reduces backbone network traffic
In many cases, we observe that a significant portion of the dataset is transferred multiple times over the network for various reasons.
To take advantage of this resue, the High-Energy Physics (HEP) community has established a number of regional storage caches~\cite{socalrepo2018, datalakes, osg}.
Analyses show that these caches could significantly reduce the data access latency as well as the traffic on the internet backbone~\cite{copps2021}.
%in-network data caching strategy includes reducing the redundant data transfers and consequently saving network traffic volume which have been described in . 

%3 in-network caching might be useful
%\fix{is there a difference between "data lake" and "in-network caching"?  -- the citations given here are talking about "data lakes". Added copps2021 for in-network caching}
%The in-network caching in the middle of a region~\cite{copps2021, datalakes, osg} would reduce the data access latency and increase the overall application performance.
In the example of the HEP community, the largest data source is the LHC instrument at CERN in Switzerland.
The main collaborations involved in generating and analyzing these data, known as ATLAS and CMS.
Their Tier-1 storage sites in the US are at Brookhaven National Laboratory and Fermi National Accelerator Laboratory respectively.
The wide-area network traffic for retrieving and replicating their data is primarily carried on the Energy Science Network (ESnet), one of the key components of the internet backbone especially designed for our nation's science and research communities.
Because the data lakes have demonstrated their effectiveness in reducing the load on the internet backbone, we are interested in exploring the predictability of their impact and the potential for providing a more general distributed storage caching strategy known as in-network caching~\cite{Li:2013:INC, Kalla:2016:INC, Sim:2022:INC}. 
%\fix{Please review above justification for doing prediction and considering in-network caching}
%Efficient management of the in-network caches is another important aspect for the resources and users, and long-term resource planning would provide sustainability for the data access. Understanding data access trends would provide insights into these challenges.

More specifically, our work starts with a study of data access trends with one of the data lakes named Southern California Petabyte Scale Cache (SoCal Repo)~\cite{socalrepo2018}. We examine the trends of network traffic volume and establish a machine learning model to predict the future network bandwidth requirement for the regional data cache.
%\fix{Are we actually making these predictions?  If not, please adjust.}
%The average network transfer traffic demand is reduced by a factor of 1.30 over the study period, while a factor of 2.35 is observed before new cache nodes have been added to the regional cache in late Aug. 2021.

The key contributions of this paper can be summarized as follows: (1) our study finds find that the SoCal Repo was able to reduce the traffic by 23\% over the study period, and by 57\% under normal usage; (2) this network traffic reduction is stable and predictable by LSTM, with 88.4\% accuracy; (3) because of the network traffic reduction, we recommend a general in-network cache to supplement the existing data lakes from HEP to benefit all science user communities.
\vspace{-0.4cm}
% where 80\% and 89\% of the true values fall into the 2 STDs and 3 STDs of the predicted values respectively
%\fix{Is it possible to review the main contributions of this work as follows: (1) our study finds the the SoCal Repo was able to reduce the traffic by 23\% over the study period; (2) this network traffic reduction is stable and predictable by LSTM, where 80\% and 89\% of the true values fall into the 2 STDs and 3 STDs of the predicted values respectively; (3) because the network traffic reduction is significant, we recommend a general in-network cache to supplement the existing data lakes from HEP to benefit all science user communities. [AS] reads well, and after Jack verifies the prediction accuracy, it's good to go.}
%\fix{[John] I'd suggest we organize the contributions of this work as: (1) study of SoCal Repo, (2) show network traffic reduction, (3) something about prediction and resource allocation for in-network caching and network resource provisioning. [AS] Correct. (1) and (2) should be in the section 3, and (3) in section 4. }

% \begin{figure*}%[htb!]
% \centering
% \includegraphics[width=0.9\textwidth]{figs/sample1.png}
% \caption{
%     Figure 1
% }
% \label{fig_access}
% \end{figure*}
\section{Background}
\label{sec:background}
Southern California Petabyte Scale Cache (SoCal Repo)~\cite{socalrepo2018} is a regional "Data Lake"~\cite{datalakes, osg} based on XCache~\cite{stashcache, xcache2014, socalrepo2018}.
XRootD system is the bases for the XCache, and supports unique capabilities for data distribution and access, especially for large collaborations such as the Large Hadron Collider (LHC)~\cite{xrootd2005, xrootdcms}.
SoCal Repo consists of 24 data cache nodes at Caltech, UCSD, and ESnet with approximately 2.5PB of storage capacity, supporting client computing jobs for High-Luminosity Large Hadron Collider (HL-LHC) analysis in Southern California.
In this cache installation, there are 11 nodes at Caltech with storage sizes ranging from 96TB to 388TB; 12 nodes at UCSD with 24TB each node; one node at an ESnet endpoint at Sunnyvale, CA with 44TB of storage.
The two southern California sites are within 200 km from each other and have a round trip time (RTT) of less than 3 milliseconds (ms) from each other, while the ESnet node is about 700 km away from UCSD, with an RTT of about 10ms. %\fix{Please check to make sure the numbers mentioned here are accurate! Ok. If slightly off, we can fix it for the final paper after verifying with UCSD/Caltech people.}
One node at Caltech is designated for NANOAOD and all other cache nodes are for MINIAOD~\cite{miniaod2019}.

% runs on a worker node at a Caltech or UCSD T2 center and
When a user's computing job needs a file from SoCal Repo, the system first looks up the location of the file using the "Trivial File Catalogue" (TFC)~\cite{Fajardo2020, Fajardo20201}.
Following the established convention for the tiered storage system, the data files are grouped into the namespace for the local cache nodes and the TFC points to a "local redirector" in XRootD where the "local redirector" knows all regional caches.
%\fix{I am still a little vague as the relationship among TFC, "local redirector", file names, physical file locations. It's ok. This is more of Frank's words by words. For the final paper, we'll have him review.}
If one of the cache nodes has the file, the redirector routes the application request to the node.
If none of them has the file, one of them is told to invoke an XRootD client to fetch the file.
The XRootD client is configured to get the file from the national XRootD data federation to the local cache node.
Local cache nodes do not connect to another cache node but always connect to the higher tier of the federation. % of the data origins.
In CMS collaboration, data federation is hierarchical where the US is one flat layer and the rest of the world is another flat layer.
%The US is connected to the rest of the world via one layer above, which means that a file is always looked for first in the US before going elsewhere.
%\fix{The current description is not very clear to me.  Is there a picture to illustrate the relationship among the layers?  If not, it might be OK to omit the details about how the US layer is related to the rest of the world.}
%In addition, one copy of all files that are part of the namespace resides in the cache nodes in the US.
%It results that cache misses should always be served from within the US, but there could be edge cases that a file possibly gets pulled from outside the US. 
By design, each file available to the CMS collaboration has at least a copy somewhere in the US.
Thus it is possible to find a copy of any file needed for analysis even though the lookup mechanism in TFC does not always guarantee to recommend a replica in the US.
%\fix{Please verify this rewrite is still accurate}

% cache replacement policy
Most of the file reads in CMS based on XRootD are vectors of byte ranges, and a cache miss leads to a vector of byte ranges getting fetched.
When new cache nodes have been added to the local cache nodes, all cache misses go to the new cache nodes first, so that the distributed cache nodes avoid deletions of old data as long as there is a new space to fill. It means that cache nodes that have been around for some time will tend to have data that is not of interest to as many users, and those data will eventually get deleted when running out of space. 
Adding more cache nodes to an already full distributed cache invariably leads to skewed distributions of data access patterns. This happened around Aug. 26, 2021 when 7 new nodes at Caltech (xrd 3-8, 11) are added to the system, and around Sep. 30, 2021 when 2 new nodes at Caltech (xrd 9-10) are added to the system. The new cache nodes get the new data. The new data is of more interest and leads to more accesses. Old data does not get deleted as there is still space on the new nodes. At some point, it will resolve itself, but may take some time to resolve.
%During the study period of 7 months from July 2021 to Jan 2022, we observed the total of 6.2 million data accesses with 5.3 PB of data transfers and 1.6 PB of data sharing.

%Southern California Repository~\cite{datalakes, osg}
%XCache~\cite{copps2021}

\section{Data Access Trends}
\label{sec:data}

Our work is based on monitoring information collected from the SoCal Repo between July 2021 and January 2022.
The collected information includes the following attributes about every data access request: user id, file id, file path, file size, the data transmission start time, the data transmission finished time, the total size of the transmission, whether the data request is a data transfer (cache miss) or data share (cache hit), which cache node the request is sent to, whether the transmission is successful, and so on.
A total of about 7.5 million data access requests are included in this study. 

%\fix{Provide a description of the raw data used for this work -- importantly, what information is available for this work: this information is needed to define the concepts such as "number of accesses", "data transferred", "data shared," and so on. @Jack, can you include important variables? Basically what's in the log? FIXED: It there anything I need to add here?}

\vspace{-0.2cm}
\begin{table}[h!]
\scriptsize
\centering
\caption{
Summary statistics for data accesses}
%\fix{Please review the numbers here: (1) the value in the row labeled Total seem to not include the Jan 2022 values, (2) "\% of Shared Data Size" might be better as "\% Shared (size)", (3) the numbers in the column labeled "\% Shared" seem to be slight off, e.g., I compute the percentage for to be 10.33\%, not 9.37\%, more significantly, the percentage for the Total is close to 30\% instead of 6.25\%.FIXED}

\begin{tabular}{|c||c|c|c|c|} \hline
  & {\# of Accesses} & \shortstack{Data Transfer \\ Size (TB)} & \shortstack{Shared Data \\ Size (TB)} & \shortstack{Net \ Traffic\\ Reduction} \tabularnewline \hline \hline
July 2021 & 1,182,717 & 385.78 & 519.25 & 57.37\%  \tabularnewline \hline
Aug 2021 & 1,078,340 & 206.94 & 313.46 & 60.23\%  \tabularnewline \hline
Sep 2021 & 1,089,292 & 206.96 & 257.18 & 55.41\%  \tabularnewline \hline
Oct 2021 & 1,058,071 & 412.18 & 141.91 & 25.61\%  \tabularnewline \hline
Nov 2021 & 878,703 & 649.30 & 82.67 & 11.29\%  \tabularnewline \hline
Dec 2021 & 983,723 & 1,257.89 & 130.03 & 9.37\%  \tabularnewline \hline
Jan 2022 & 1,207,332 & 2,238.59 & 148.26 & 6.21\%  \tabularnewline \hline \hline
Total &  7,478,178 & 5,357.67 & 1,592.79 & 22.91\%  \tabularnewline \hline
Daily Average & 35,441.60 & 25.51 & 7.55 & 22.83\%  \tabularnewline \hline
\end{tabular}
\label{tab:summary_data_all}
\end{table}

Table~\ref{tab:summary_data_all} shows the basic statistics about the data accesses to all cache nodes during the study period (from July 2021 to Jan. 2022).
%The number of accesses counts all user requests to read a file, vector read (readv), and open a file. \fix{is this definition correct?  anything else needs to be included here? [AS] This is not correct as the data is not from xrootd logs, but from a separate monitoring service. Basically, one record is one access, either cache miss or cache hit.}
If an "access" could be satisfied with a file in a cache, then it is a cache hit.
%On the other hand, if the requested file needs to be retrieved from another "datalake" or another tier of the storage federation, then it is a cache miss.
On the other hand, if the requested file needs to be retrieved from a remote storage site, then it is a cache miss.
Cache miss would require a data file to be transferred from a remote site over the wide-area network.
%Each data access is a cache hit or a cache miss depending on whether the data file is already in the cache when the request arrives.
The "Data Transfer Size" in the table is the total volume of data transferred to satisfy the cache misses.
%when a data file was accessed for the first time. For cache misses, the caching nodes in the region did not have the data, resulting in a data transfer from the remote data source to one of the local cache nodes.
The "Shared Data Size" refers to the total volume from the cache hits.
%, when the data file was already in the cache and readily available for the application to access. The shared Data access count is for repeated accesses only, and corresponds to the network traffic savings.
The "Net Traffic Reduction" is the percentage of network traffic reduction by the cache system, calculated monthly by (shared data size) / (total access size). 

% The "Percentage of Shared Data" is the ratio between the shared data and the transferred data displayed as percentages.

%\fix{We might not be able to do much about this concern, but from earlier studies, we see that 2/3 of data accesses are cache misses, which should translates to Percentage of Shared Data to be about 200\%.  My thinking is that we should re-examine how we count the data accesses;-)  [AS] Shared Data is cache hits. Let's think about this after the submission. Next Thursday we can discuss this with Jack.}
%column indicates the percentage of the total data access size that was shared rather than transferred. This percentage is closely related to the network traffic demand reduction. 

Table~\ref{tab:summary_data_all} shows the net traffic reduction was about 60\% during the first three months of the observation, but dropped to as low as 6\% in January 2022.
This drop is due to a usage change among the physicists in the region, as some users are streaming data through the caching system. 
% because new cache nodes have been added every month since Sep. 2021, and the design of the regional cache directs the cache misses to new nodes. 

\begin{figure}[H]
  \includegraphics[width=\linewidth]{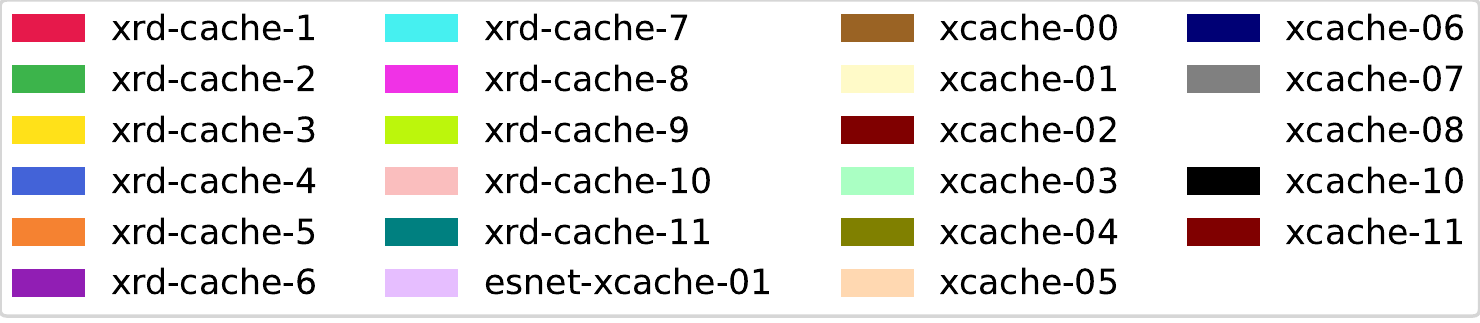}
  \caption{Legend for each node in the regional cache}
  \label{fig:legend}
\end{figure}
Figure \ref{fig:legend} indicates the color for each node in all the following plots unless specified otherwise. 

The monitoring system had troubles on Nov. 24, 2021, and from Dec. 15, 2021 to Dec. 18, 2021. So there are no data during these periods, showing gaps in the following daily plots during these periods.
%\fix{It might be more related to usage pattern change than deployment of new cache nodes -- is there a way to verify this? [AS] This is not of the new cache node effect. Either monitoring issue or usage pattern. No way to verify this.}
%%%%%%%%%%%%%%%%%%%%%%%%%%%%%%%%%%%%%%%%%%%%%%%%%%%%%%%%%%%%%%%%%%%%%%
% Access Count
%\subsection{Data Access Count}

\begin{figure}[htb!]
\centering
\subfloat[Daily]{%
      \includegraphics[width=\linewidth, height=2.5cm]{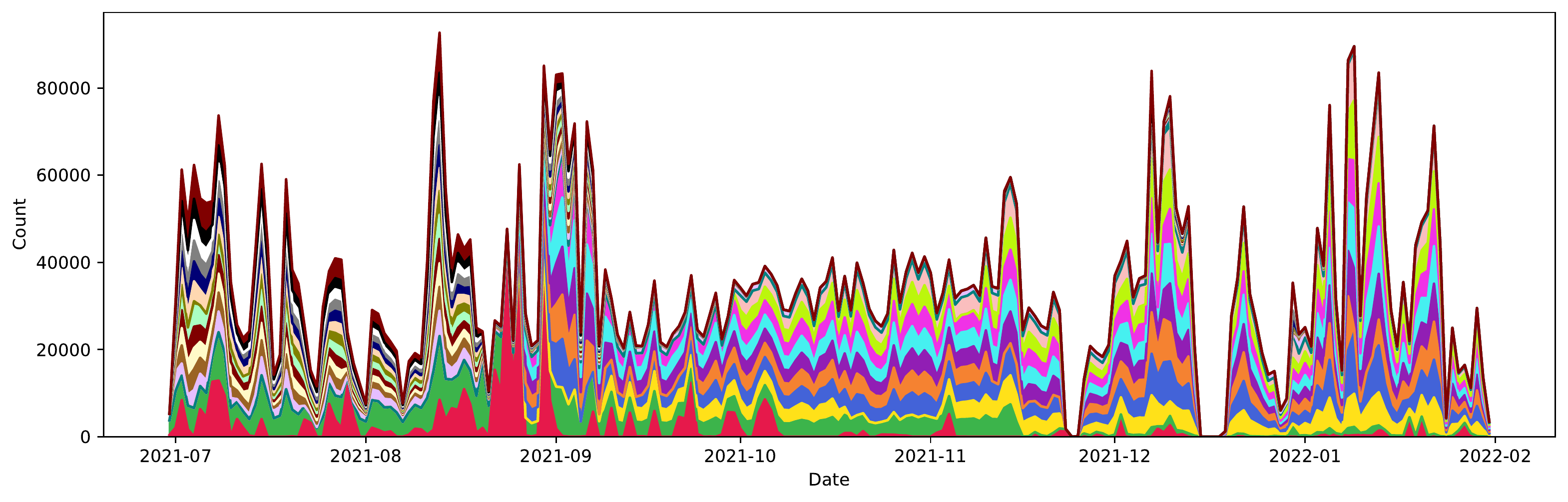}
      \label{fig:access_count}
} \vspace{-0.2cm} \newline
\subfloat[Weekly]{
  \includegraphics[width=\linewidth, height=2.5cm]{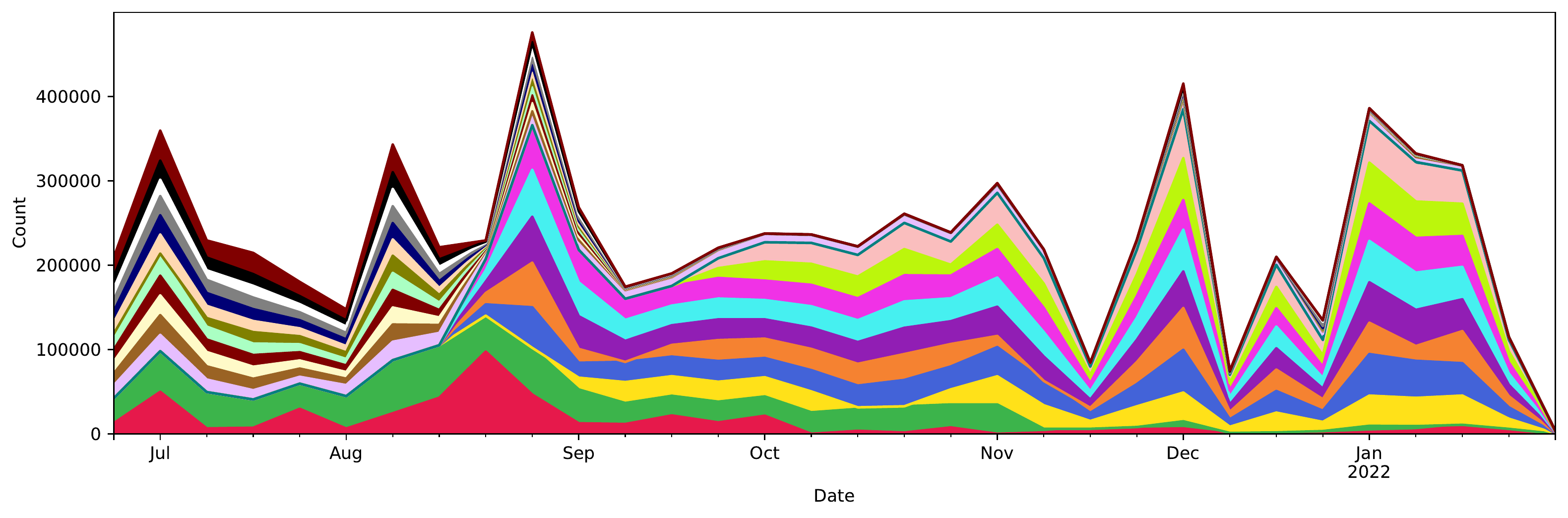}
  \label{fig:weekly_access_count}
}
\caption{Total data access counts in the regional cache.  The number of access is relatively stable during the time period of this study.}
\vspace{-0.5cm}
\end{figure}

Figure \ref{fig:access_count} shows the daily total data access counts, combining the number of date shares (i.e. cache hits) and data transfers (i.e. cache misses), and the distribution among the cache nodes.  Figure \ref{fig:weekly_access_count} shows the weekly total data access counts and distribution among the cache nodes. 
%\fix{mention the dates to two gaps in Figure \ref{fig:access_count}. FIXED,THE EXACT DATES ARE PUT RIGHT BELOW THE LEGEND}

The number of total accesses is fairly consistent throughout the study period, fluctuating around 31,000 per day. Each cache node evenly receives file requests before September 2021. When new cache nodes have been added to the regional cache, many of the cache accesses have been sent to the new cache nodes evenly with the previously described reason in Section \ref{sec:background}.

\iffalse
\begin{figure}[H]
  \includegraphics[width=\linewidth]{figs/plots/myplots/access_count_report.pdf}
  \caption{Daily total data access counts of total accesses on each node in the regional cache}
  \label{fig:access_count}
\end{figure}

\begin{figure}[H]
  \includegraphics[width=\linewidth]{figs/plots/myplots/weekly_access_count_report.pdf}
  \caption{Weekly total data access counts of total access on each node in the regional cache}
  \label{fig:weekly_access_count}
\end{figure}

 \begin{figure}[H]
  \includegraphics[width=\linewidth]{figs/plots/myplots/access_count_xrd1_report.pdf}
  \caption{Daily data access count of Xrd1}
  \label{fig:xrd1_access_count}
\end{figure}

\begin{figure}[H]
  \includegraphics[width=\linewidth]{figs/plots/myplots/access_count_noxrd1_report.pdf}
  \caption{Daily total data access counts of total accesses on each node in the regional cache without Xrd1}
  \label{fig:noxrd1_access_count}
\end{figure}

\fi

%%%%%%%%%%%%%%%%%%%%%%%%%%%%%%%%%%%%%%%%%%%%%%%%%%
% Access Size
%\subsection{Data Access Size}

\begin{figure}%[htb!]
%\fix{Is it possible to have the two figure's x-axis the same? MERGING THESE TWO IS VERY HARD. THE XTICKS DOES NOT ALIGN: ONE IS DAILY ONE IS WEEKLY}
\centering 
\subfloat[Daily]{
  \includegraphics[width=\linewidth, height=2.5cm]{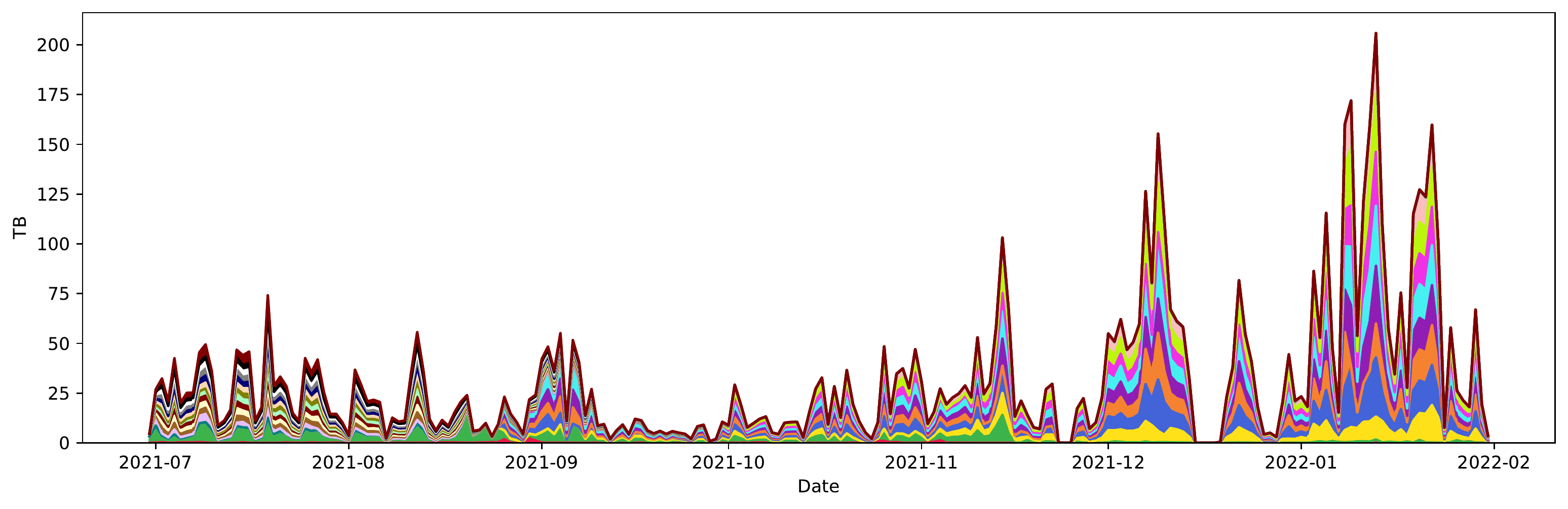}
  \label{fig:access_size}
} \vspace{-0.2cm} \newline
\subfloat[Weekly]{
  \includegraphics[width=\linewidth, height=2.5cm]{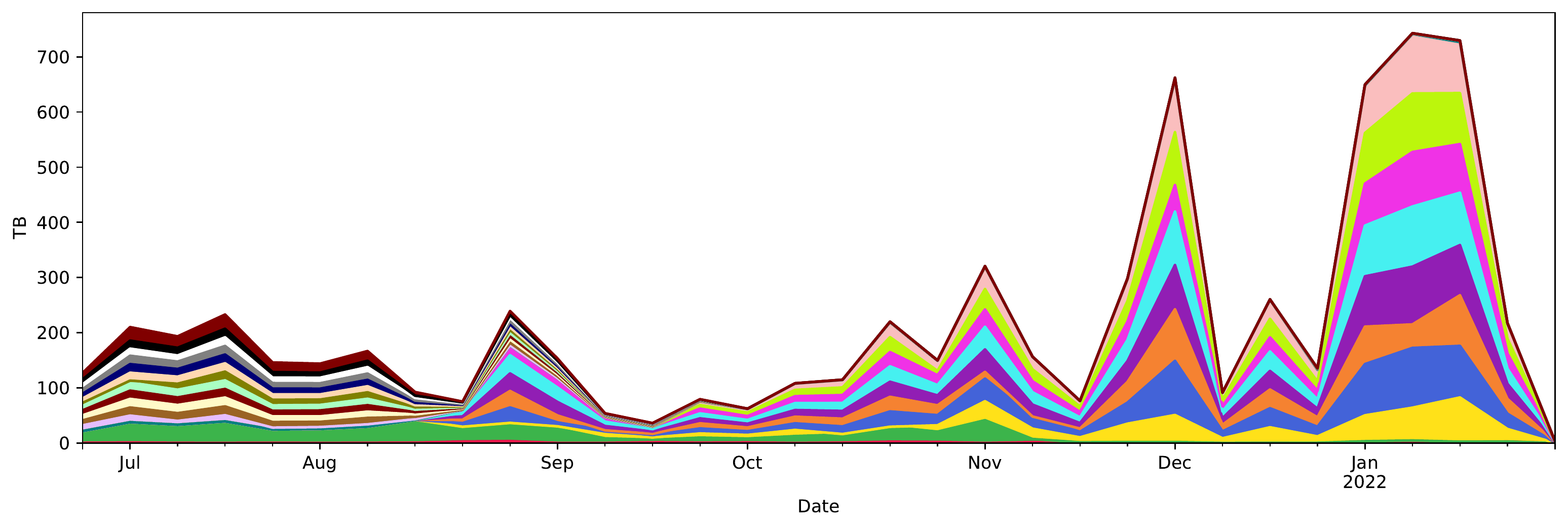}
  \label{fig:weekly_access_size}
} \newline
    \caption{Total data access sizes in the regional cache}
\vspace{-0.4cm}
\end{figure}

Figure \ref{fig:access_size} shows the daily total data access sizes, combining shared data sizes (i.e. cache hits) and transferred data sizes (i.e. cache misses) on each cache node.
Figure \ref{fig:weekly_access_size} shows the weekly total data access sizes among cache nodes.
The total access size is increasing over the study periods indicating that the requested data size grows while the number of accesses remains about the same each month. When traffic is relatively small, the daily traffic volume is about 21TB per day. After new cache nodes have been added to the regional cache, many of the data access traffic have been sent to the new cache nodes, and it is expected by the policy described in Section \ref{sec:background}.

\iffalse
\begin{figure}[H]
  \includegraphics[width=\linewidth]{figs/plots/myplots/access_size_report.pdf}
  \caption{Daily total data access sizes of total access sizes on each node in the regional cache}
  \label{fig:access_size}
\end{figure}

\begin{figure}H[]
  \includegraphics[width=\linewidth]{figs/plots/myplots/weekly_access_size_report.pdf}
  \caption{Weekly total data access sizes of total access sizes on each node in the regional cache}
  \label{fig:weekly_access_size}
\end{figure}

\begin{figure}[H]
  \includegraphics[width=\linewidth]{figs/plots/myplots/access_size_xrd1_report.pdf}
  \caption{Daily data access size of Xrd1}
  \label{fig:xrd1_access_size}
\end{figure}

\begin{figure}[H]
\includegraphics[width=\linewidth]{figs/plots/myplots/access_size_noxrd1_report.pdf}
\caption{Daily total data access counts of total accesses on each node in the regional cache without Xrd1}
\label{fig:noxrd1_access_size}
\end{figure}

\fi

%%%%%%%%%%%%%%%%%%%%%%%%%%%%%%%%%%%%%%%%%%%%%%%%%%%%%%

%\subsection{Average Size per Access}

% Average Size per Access
\begin{figure}%[htb!]
%\fix{It is possible to have only the average line (i.e., the bottom half) with the individual cache lines? FIXED}
\centering 
\subfloat[Daily]{%
  \includegraphics[width=\linewidth, height=2.5cm]{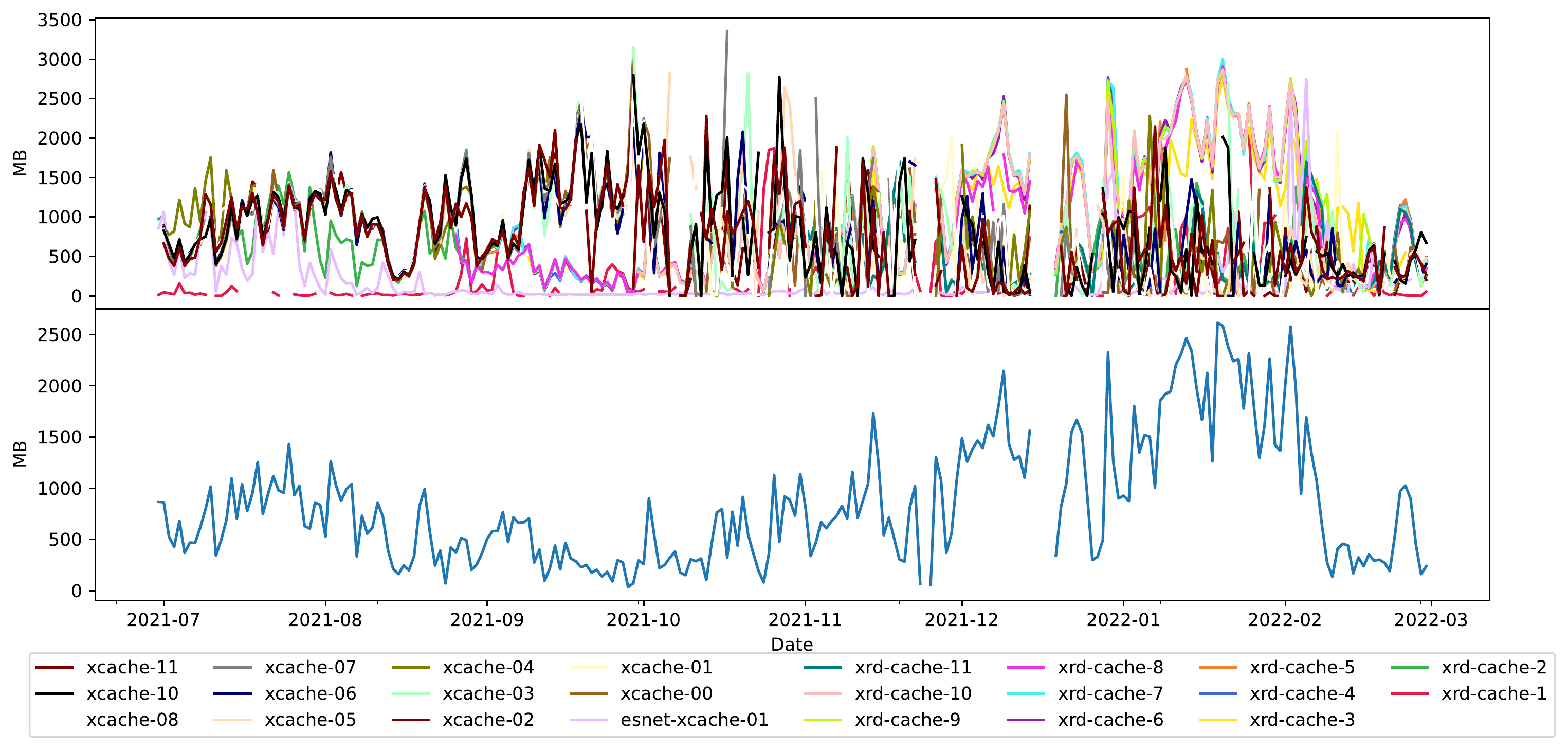}
  \label{fig:avg_access_size}
} \vspace{-0.2cm} \newline
\subfloat[weekly]{%
  \includegraphics[width=\linewidth, height=2.5cm]{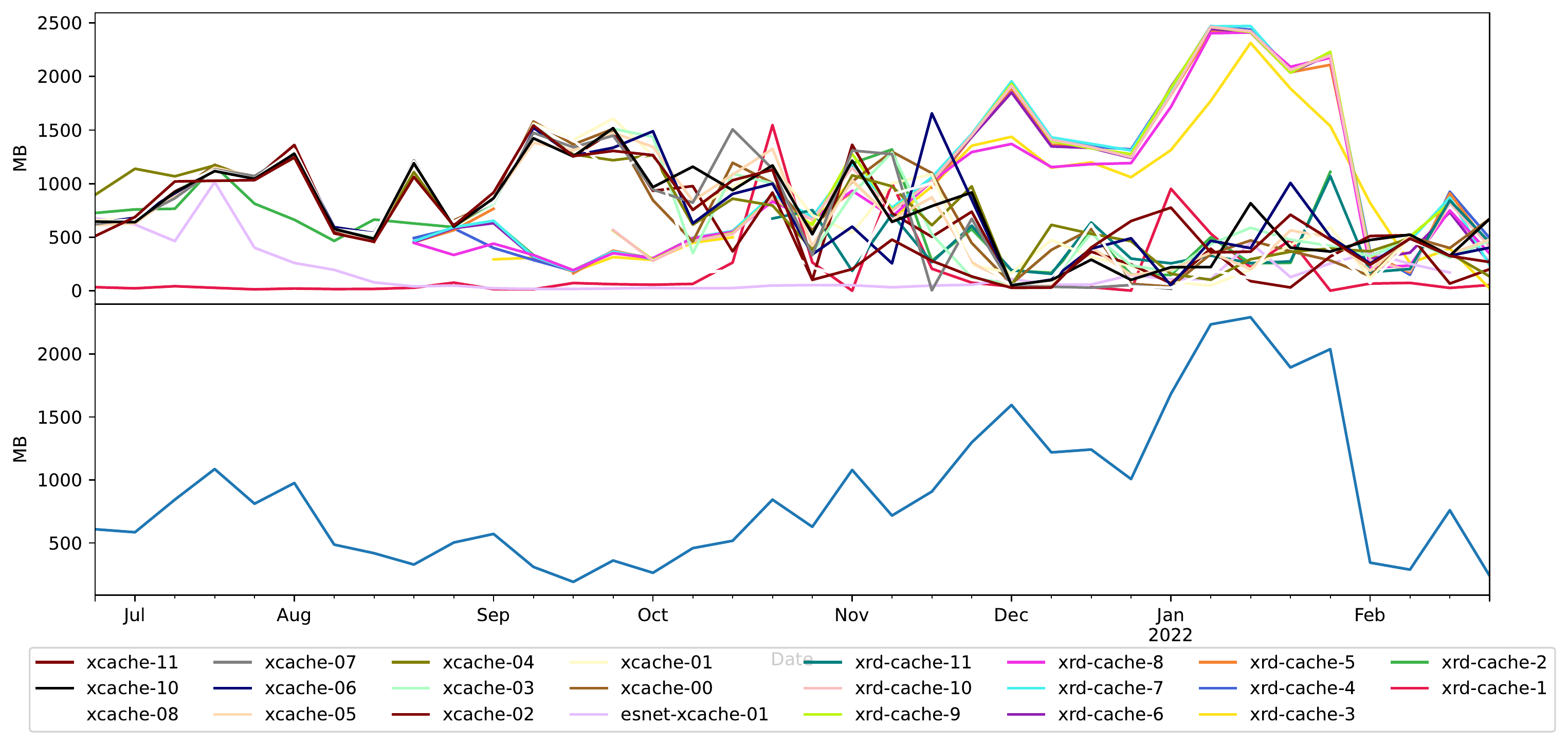}
  \label{fig:weekly_avg_access_size}
} \newline
\caption{Average data size per access in the regional cache}
\vspace{-0.5cm}
\end{figure}

Figure \ref{fig:avg_access_size} shows the average data sizes per access, calculated daily by ${(Total \ Data \ Access \ Size)}/{(Total \ Data \ Access \ Counts)}$.
Figure \ref{fig:weekly_avg_access_size} shows the weekly average data sizes per access. The upper parts of both daily and weekly plots show the average data size per access for each node, and the lower parts show the average data size per access of all nodes combined. The average data size per access of all cache nodes gradually decreases since Nov 2021. %, and the average data size per access of Xrd nodes gradually increases though out the study period. 
Overall, the average data size per access is increasing during the study period, consistent with the increases in the total access size while the data access counts remain about the same each month.

\begin{figure}%[htb!]
\centering 
\subfloat[Daily]{%
  \includegraphics[width=\linewidth, height=2.5cm]{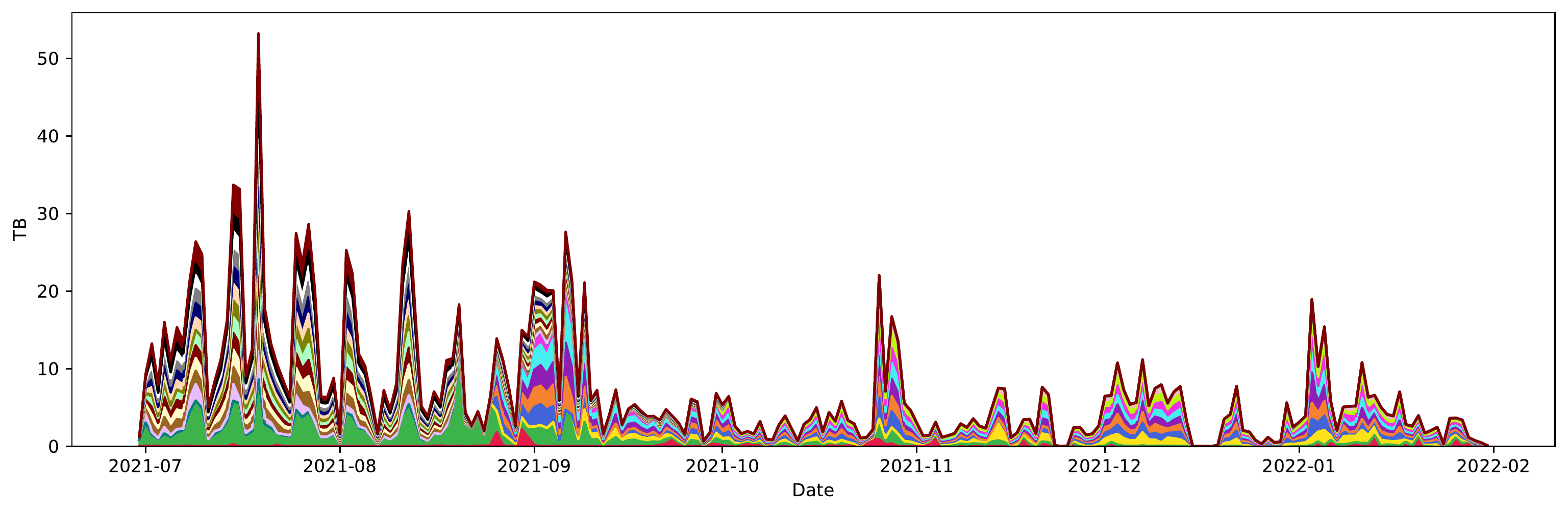}
  \label{fig:hit_size}
} \vspace{-0.2cm} \newline
\subfloat[Weekly]{%
  \includegraphics[width=\linewidth, height=2.5cm]{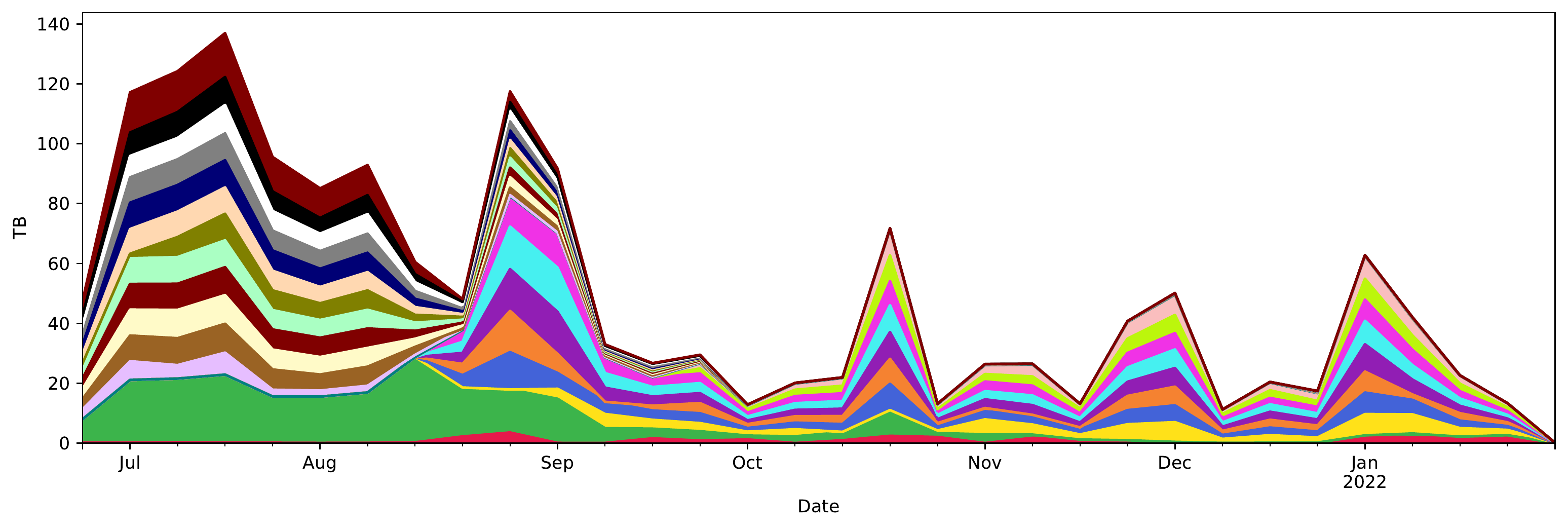}
  \label{fig:weekly_hit_size}
} \newline
\caption{Total sizes of the cache hits in the regional cache}
\vspace{-0.5cm}
\end{figure}

Figure \ref{fig:hit_size} and \ref{fig:weekly_hit_size} show the daily and weekly total shared data sizes among the cache node respectively.
The total shared data size shows a big drop since mid Sept. 2021, with only a few occasional hikes. After new cache nodes have been added to the regional cache, most of the cache hits have been sent to new cache nodes as the new nodes have recent data of more interest. 
\begin{figure}[htb!]
  \includegraphics[width=\linewidth, height=2.5cm]{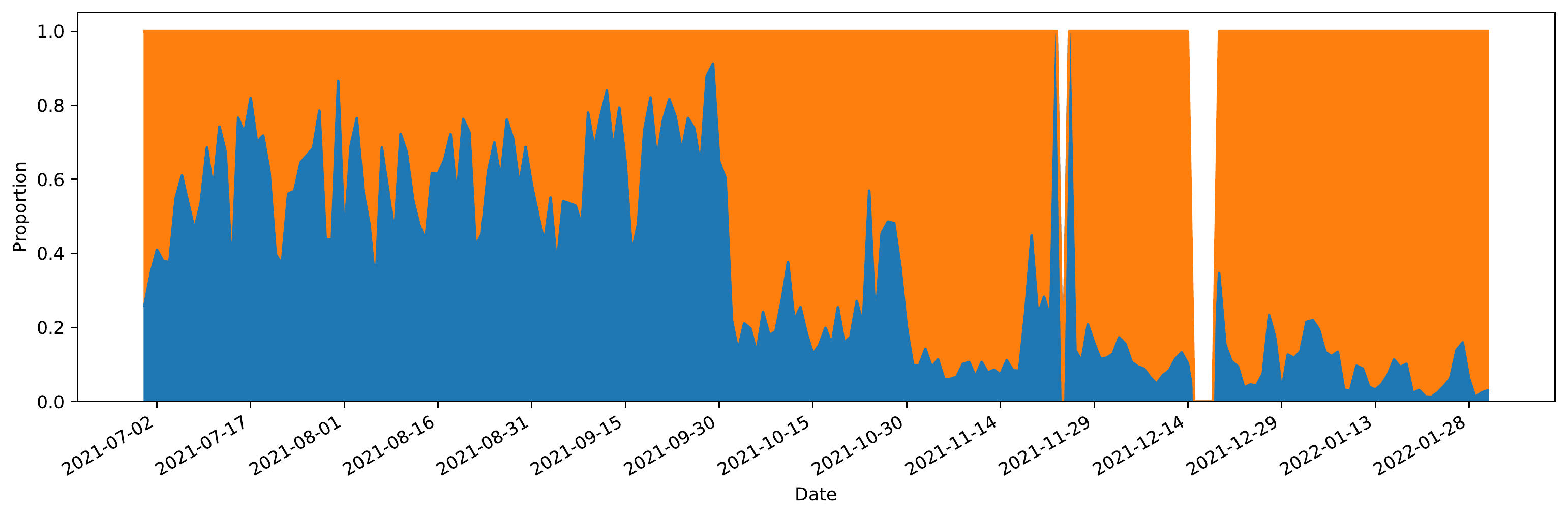}
  \caption{Daily proportion of cache miss sizes (orange area) and cache hit sizes (blue area) in the regional cache. The cache hit rate reduces after October 2021 because of a usage pattern change.}
  \label{fig:hit_size_ratio}
\end{figure}

Figure \ref{fig:hit_size_ratio} shows the proportion of the daily total  data size of the cache misses. %It is consistent to the Figures \ref{fig:hit_size} and \ref{fig:miss_size}. 
The sudden drop in the daily proportion of cache hit sizes and the gradually increasing cache miss sizes are due to changes in the access trend that several users are constantly streaming data.

\iffalse
Figure \ref{fig:hit_size_ratio} shows the proportion of the total daily size of data shared data accesses, calculated by the equation \ref{eqn:share-to-transfer}.
\begin{equation}
    Proportion \ of \ Cache \ Hit = \frac{Number \ of \ Cache \ Hits}{Number \ of \ Accesses} 
    \label{eqn:share-to-transfer}
\end{equation}

\begin{figure}[H]
  \includegraphics[width=\linewidth]{figs/plots/myplots/share_ratio_noxrd1.pdf}
  \caption{Daily proportion of data transferred sizes (orange area) and data shared sizes (blue area) in the regional cache without Xrd1}
  \label{fig:noxrd1_hit_size_ratio}
\end{figure}

Figure \ref{fig:hit_size_ratio} shows the proportion of the total daily size of data shared data accesses without counting the Xrd1 node. Xrd1 does not significantly affect the proportion of data shared data access.
\fi

%%%%%%%%%%%%%%%%%%%%%%%%%%%%%%%%%%%%%%%%%%%%%%%%%%%%%%%%%%%%%%%%%%%%%%%%
%\iffalse
% Traffic Reduction
%\subsection{Network Traffic Demand Reduction Rate}

%\begin{figure}[H]
\begin{figure}j%[htb!]
  \includegraphics[width=\linewidth]{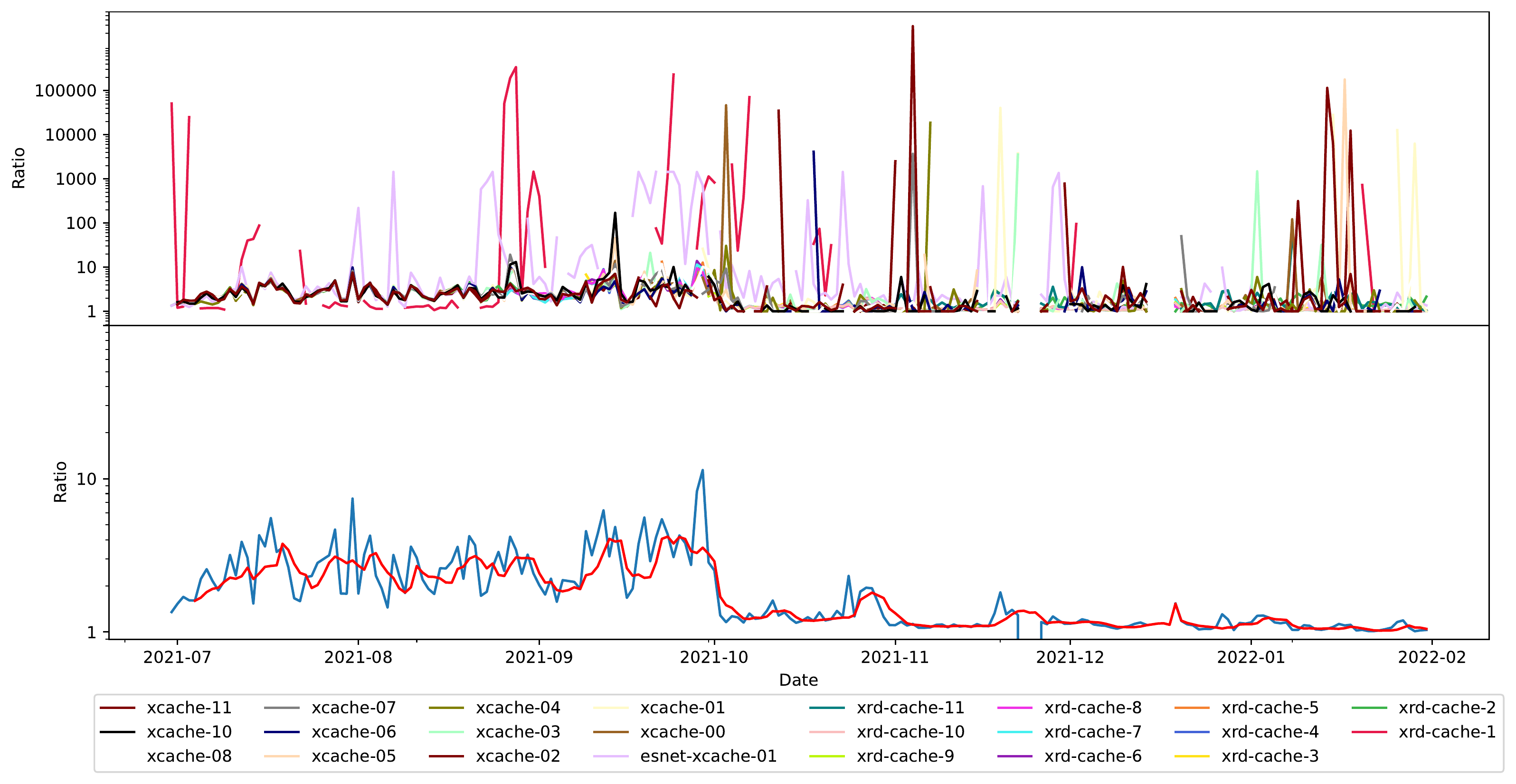}
  \caption{Daily network traffic demand reduction.  The average reduction rate is 2.35 before October 2021 and 1.11 afterward.}
  \label{fig:size_reduction}
  \vspace{-0.5cm}
\end{figure}

%Figure \ref{fig:size_reduction} shows the daily network traffic demand reduction rate for each node in the regional cache, with the red line indicating the 7-day moving average of the network traffic demand reduction rate. 
The network traffic demand reduction rates, calculated by the eqn. (\ref{eqn_traffic_reduction}), are shown in Figure \ref{fig:size_reduction} with the red line indicating the 7-day moving average of the network traffic demand reduction rate. 
The traffic demand reduction rate is the ratio of data volume users access and the volume transferred over the backbone network.
It shows that the network traffic demand reduction rate experiences a sudden drop since Oct. 2021 when the user access trends changes to streaming many new data files.
The average network traffic demand reduction rate is 1.30 during the study period, while the average rate from July 2021 to Sep. 2021 is 2.35 before the user access trends change. The average rate drops to 1.11 from Oct. 2021 to Jan. 2021, as user streaming data have a great negative impact on the statistics of the caching system.

\vspace{-0.4cm}
\begin{multline}
\text{network traffic demand reduction rate} = \\
\frac{\text{(total cache hit size + total cache miss size)}}{\text{(total cache miss size)}}
\label{eqn_traffic_reduction}
\end{multline}

\begin{figure}[htb!]
\centering 
\subfloat[Daily total data reuse size]{%
  \includegraphics[width=\linewidth]{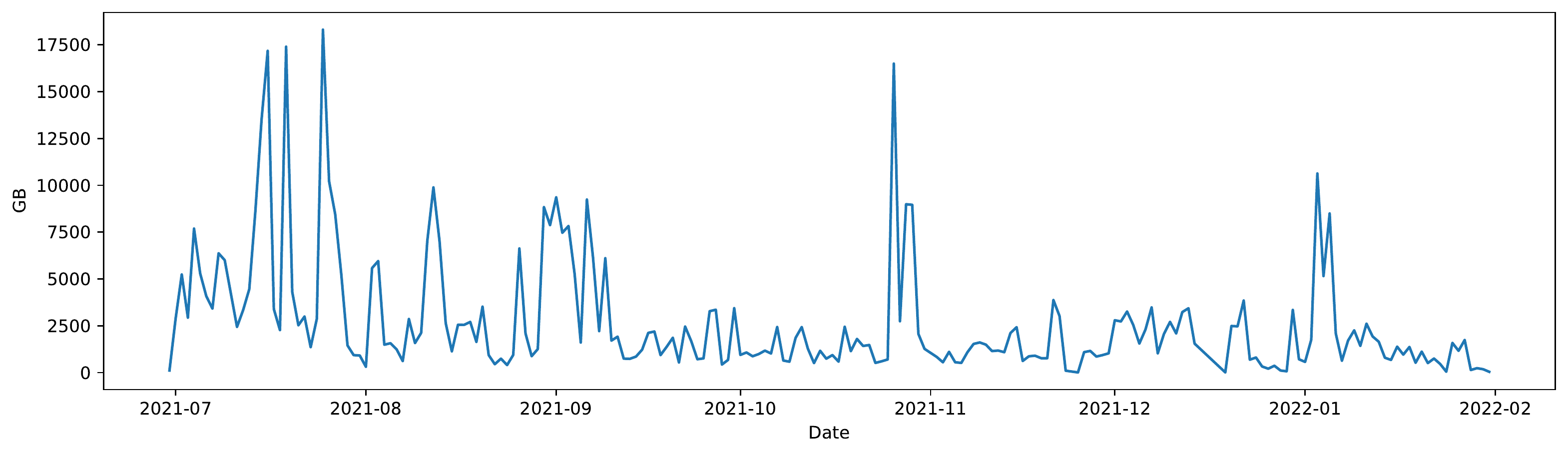}
  \label{fig:reuse_size_1}
} \vspace{-0.2cm} \newline
\subfloat[Daily total data reuse size with 7-day moving average]{%
  \includegraphics[width=\linewidth]{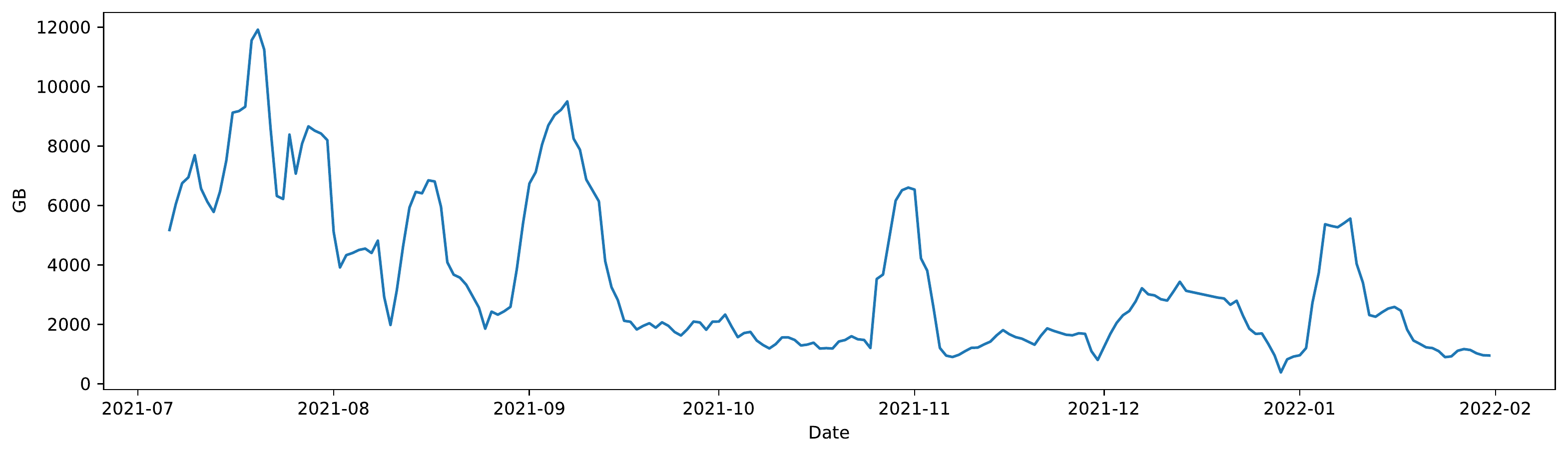}
  \label{fig:reuse_size_7}
} \newline
\caption{Daily total data reuse size in the regional cache}
\end{figure}

Figure \ref{fig:reuse_size_1} shows the daily total data reuse size for all nodes in the regional cache. Data reuse means the re-access of the same data file without transferring within the same day (i.e. successive cache hits on the same data without a cache miss on that data during one day. 
Data reuse indicates the network traffic savings on files that are accessed multiple times. The total data reuse size is the total size of data reused in a single day. 
Figure \ref{fig:reuse_size_7} shows the daily total data reuse size of 7-day moving average in the regional cache.
Prior to Oct. 2021 before the user behavior changes, the total data reuse size generally follows the access size. Since then, the  total data reuse size is relatively stable with a few spikes in the middle.

\begin{figure}[htb!]
\centering 
\subfloat[Daily data reuse rate]{%
  \includegraphics[width=\linewidth]{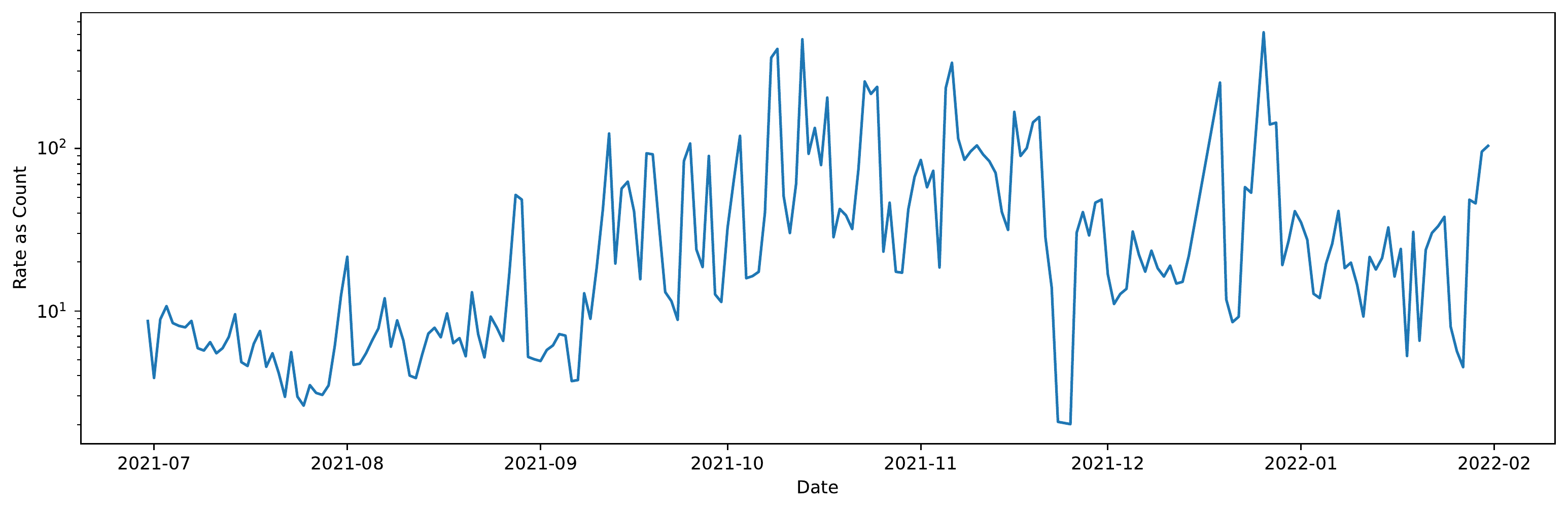}
  \label{fig:reuse_rate_1}
} \vspace{-0.2cm} \newline
\subfloat[Daily data reuse rate with 7-day moving average]{%
  \includegraphics[width=\linewidth]{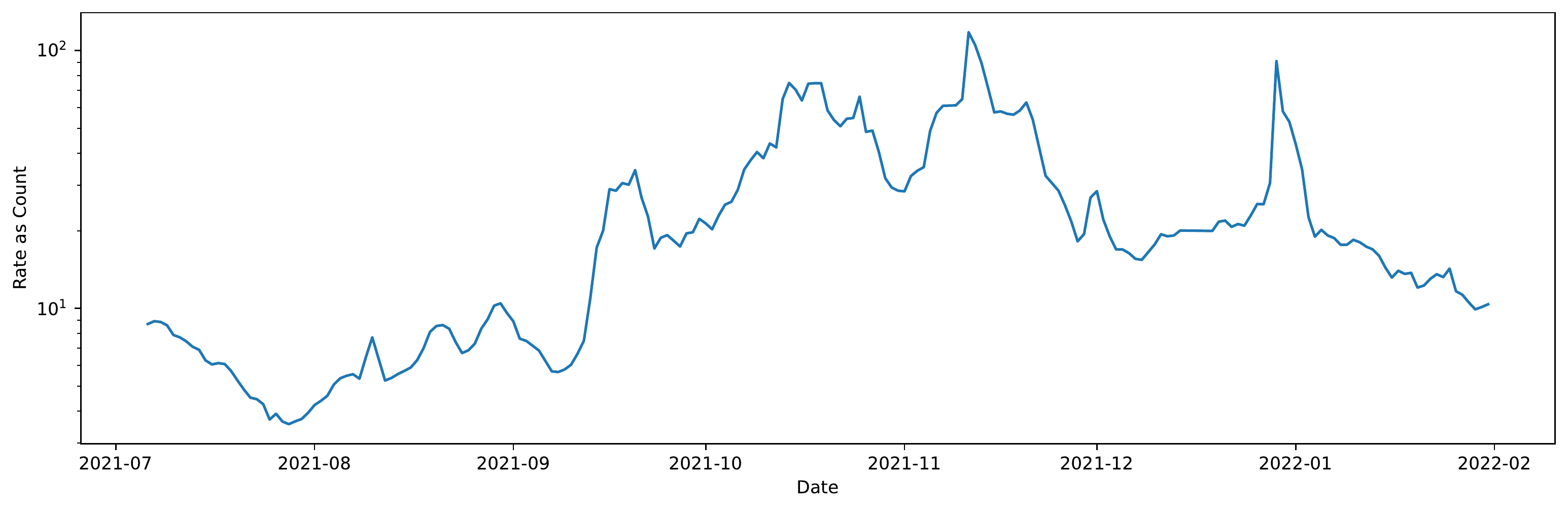}
  \label{fig:reuse_rate_7}
} \newline
\caption{Daily data reuse rat.  Despite the significant volume of data being streamed through this cache, there are still significant amount of reuse of files in cache.}
\vspace{-0.3cm}
\end{figure}

Figure \ref{fig:reuse_rate_1} shows the daily data reuse rates for all nodes in the regional cache. The data reuse rate is the number of times that the files have been reused in a single day, calculated by \\
${(Total \ Data \ Reused \ Count)} / {(Number \ of \ Unique \ Reused \ Files)}$. \\
Figure \ref{fig:reuse_rate_7} shows the daily data reuse rate of the 7-day moving average for all nodes in the regional cache. 
It’s measuring how well the caching system saves the traffic on files that are accessed multiple times. The daily data reuse rate increases gradually from July 2021 to mid Nov. 2021, and decreases a bit since then. The daily data reuse is not affected much by the behavior changes of several users' streaming data.

\section{Modeling and Predicting Cache Utilization}
\label{sec:eval}
To further understand the trends of cache utilization and explore the potential effectiveness of a more general caching mechanism in addition to the dedicated caching system for the specific user community, we next attempt to build machine learning models to investigate the predictability of common cache utilization trends.
We model these cache utilization measures as a time series and plan to employ a well-established recurrent neural network (RNN)~\cite{sherstinsky2020fundamentals}.
More specifically, we use a version of RNN known as Long-Short Term Memory (LSTM) in this work~\cite{sherstinsky2020fundamentals, greff2016lstm}.
%Models are built separately on the daily data and daily data with 7-day moving average.

\iffalse
\begin{figure*}%[htb!]
\centering 
\subfloat[]{%
  \includegraphics[width=0.24\linewidth]{figs/plots/mlplot/data_dist/hourDist_access_count.pdf}
  \label{subfigure: hourdist_access_count}
}
\subfloat[]{%
  \includegraphics[width=0.24\linewidth]{figs/plots/mlplot/data_dist/hourDist_access_size.pdf}
  \label{subfigure: hourdist_access_size}
}
\subfloat[]{%
  \includegraphics[width=0.24\linewidth]{figs/plots/mlplot/data_dist/hourDist_hit_count.pdf}
  \label{subfigure: hourdist_hit_count}
}
\subfloat[]{%
  \includegraphics[width=0.24\linewidth]{figs/plots/mlplot/data_dist/hourDist_hit_size.pdf}
  \label{subfigure: hourdist_hit_size}
}\newline
\subfloat[]{%
  \includegraphics[width=0.24\linewidth]{figs/plots/mlplot/data_dist/hourDist_miss_count.pdf}
  \label{subfigure: hourdist_miss_count}
}
\subfloat[]{%
  \includegraphics[width=0.24\linewidth]{figs/plots/mlplot/data_dist/hourDist_miss_size.pdf}
  \label{subfigure: hourdist_miss_size}
}
\subfloat[]{%
  \includegraphics[width=0.24\linewidth]{figs/plots/mlplot/data_dist/hourDist_reuse_count.pdf}
  \label{subfigure: hourdist_reuse_count}
}
\subfloat[]{%
  \includegraphics[width=0.24\linewidth]{figs/plots/mlplot/data_dist/hourDist_reuse_size.pdf}
  \label{subfigure: hourdist_reuse_size}
}\newline
\caption{Distribution of Hourly (a) Access Count (b) Access Size (c) Data Shared Count (d) Shared Size (e) Data Transferred Count (f) Data Transferred Size (g) Data Reuse Count (h) Reuse Size}
\label{fig:hourDist}
\end{figure*}
\fi

\begin{figure*}%[htb!]
\centering 
\subfloat[Access counts]{%
  \includegraphics[width=0.24\linewidth, height=2.5cm]{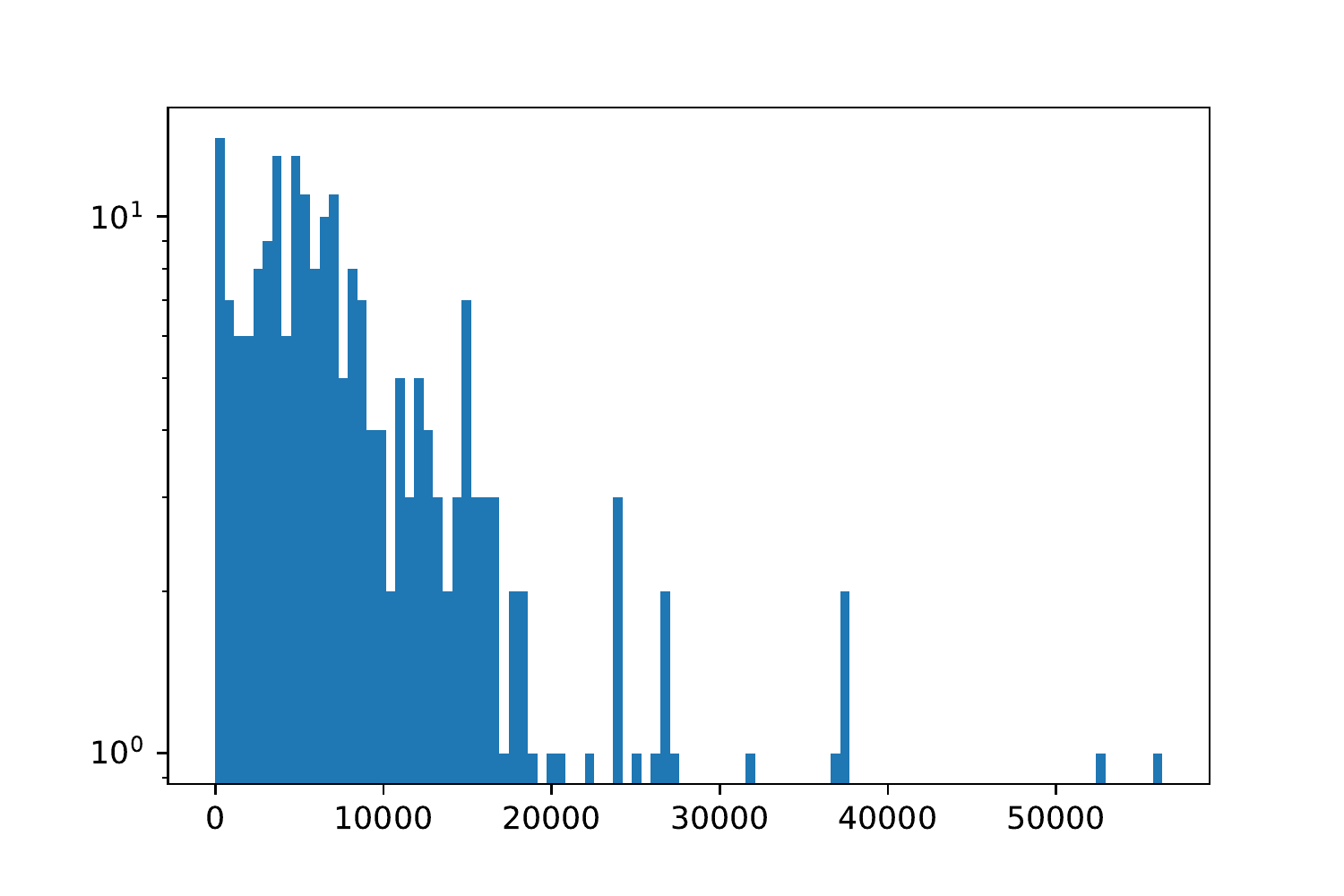}
  \label{subfigure: daydist_access_count}
}
\subfloat[Access sizes]{%
  \includegraphics[width=0.24\linewidth, height=2.5cm]{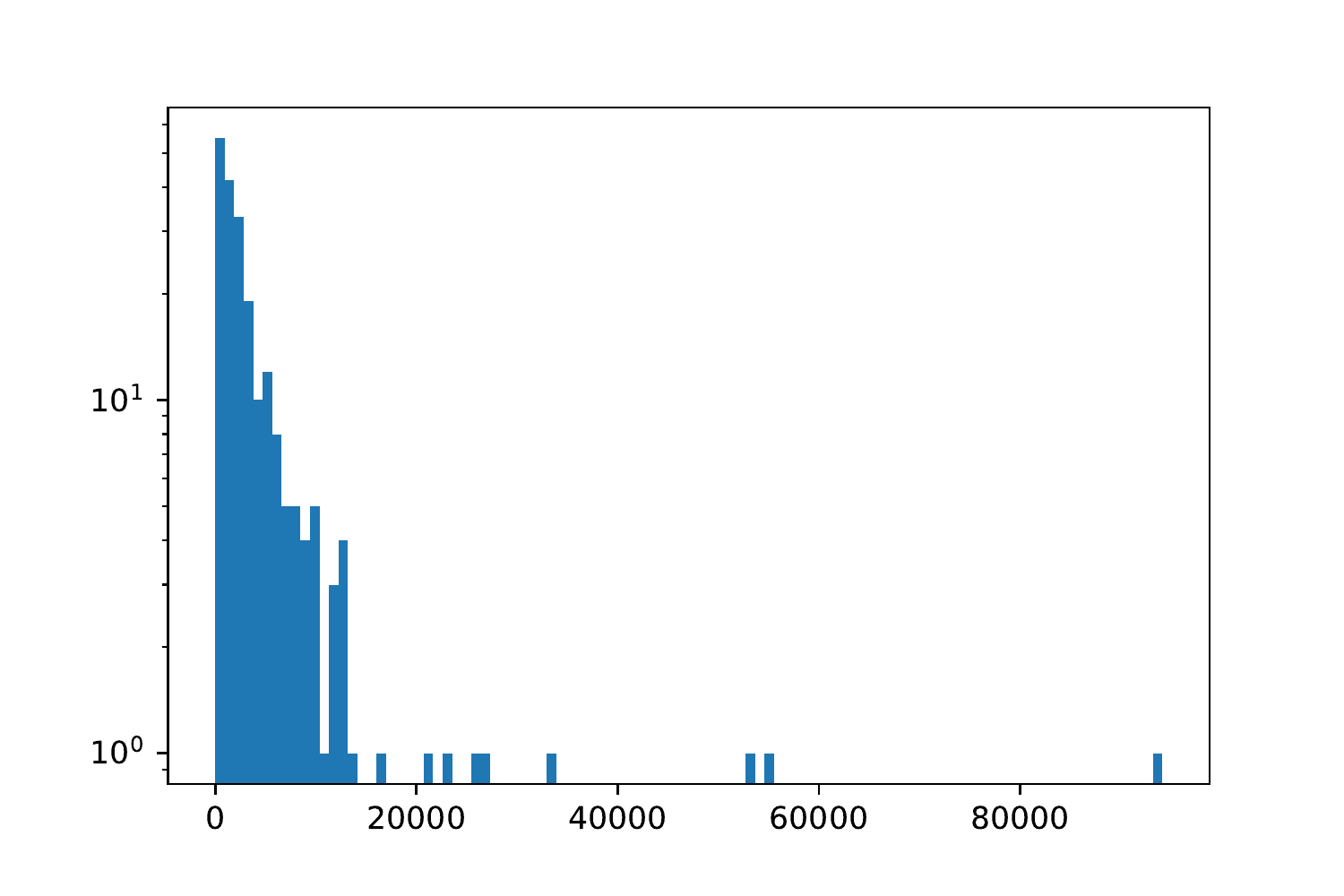}
  \label{subfigure: daydist_access_size}
}
\subfloat[Cache hit counts]{%
  \includegraphics[width=0.24\linewidth, height=2.5cm]{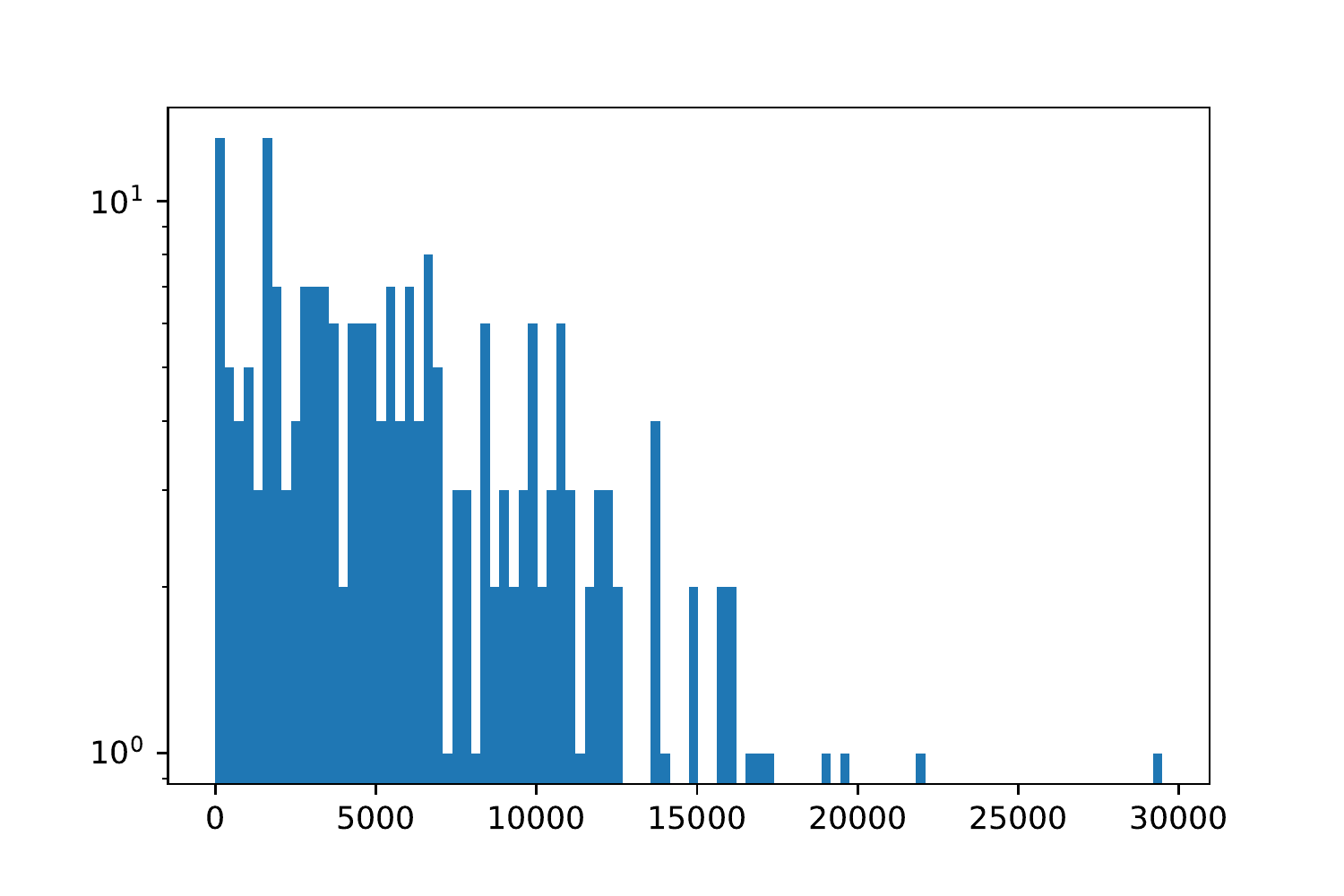}
  \label{subfigure: daydist_hit_count}
}
\subfloat[Cache hit sizes]{%
  \includegraphics[width=0.24\linewidth, height=2.5cm]{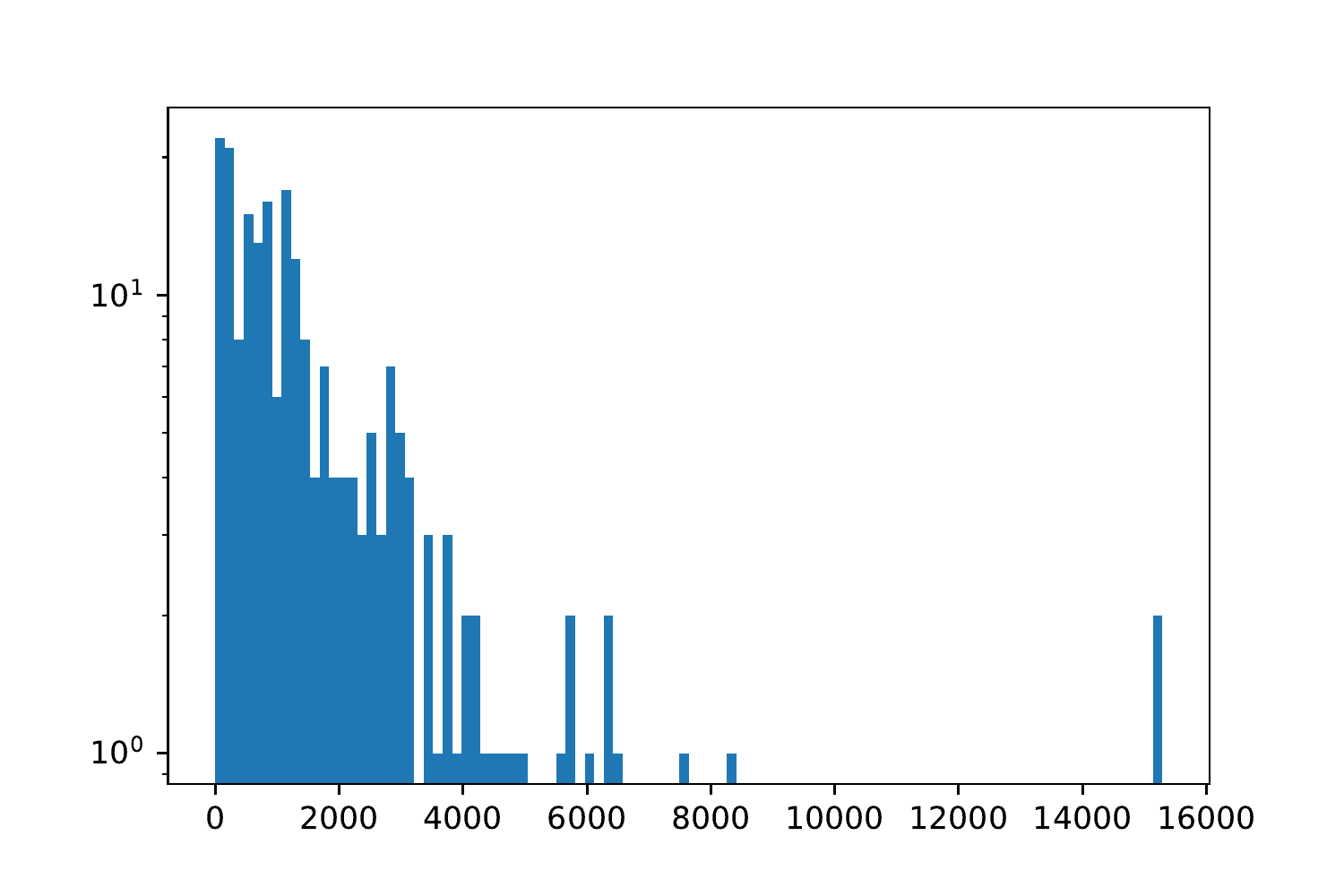}
  \label{subfigure: daydist_hit_size}
} \vspace{-0.5cm} \newline
\subfloat[Cache miss counts]{%
  \includegraphics[width=0.24\linewidth, height=2.5cm]{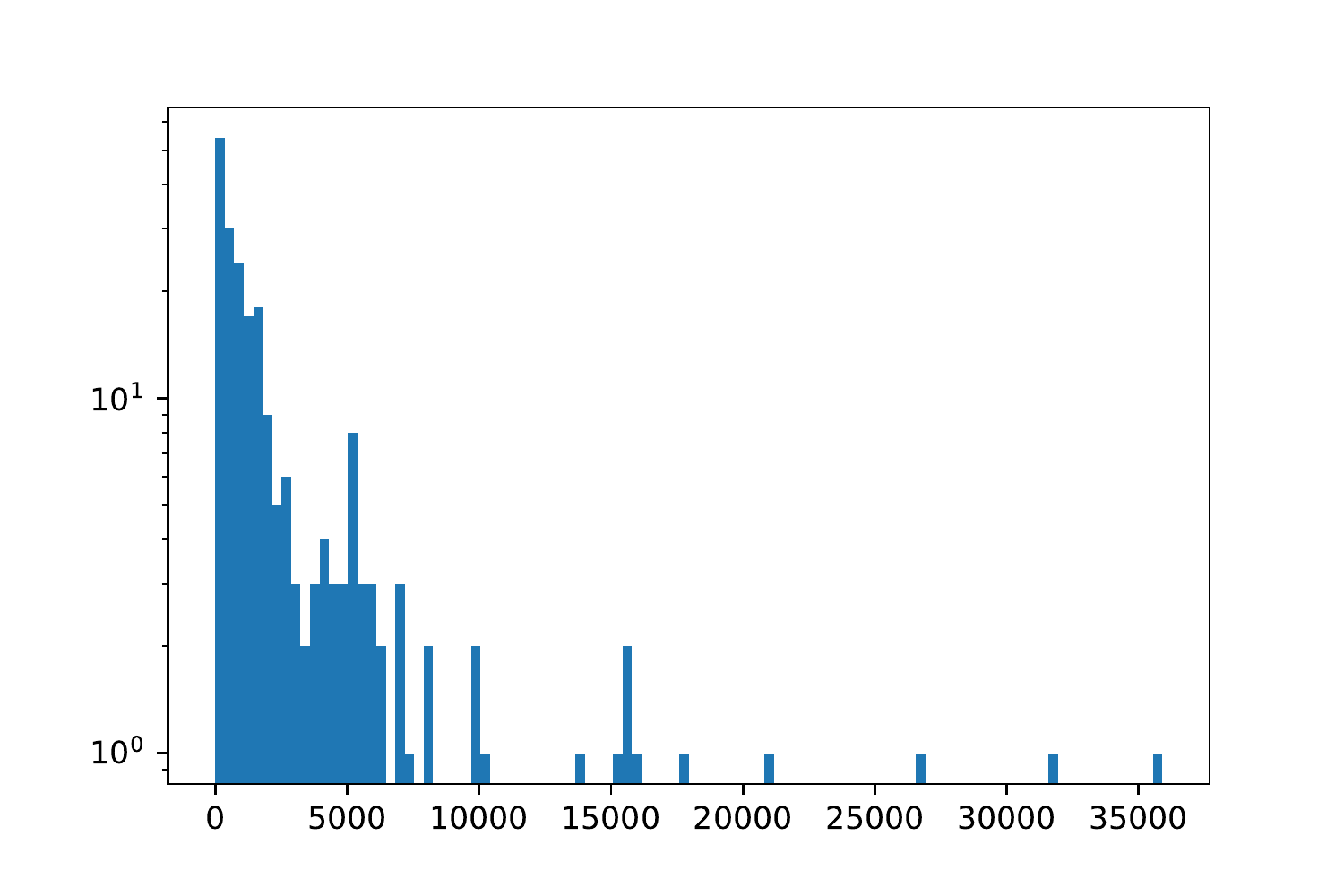}
  \label{subfigure: daydist_miss_count}
}
\subfloat[Cache miss sizes]{%
  \includegraphics[width=0.24\linewidth, height=2.5cm]{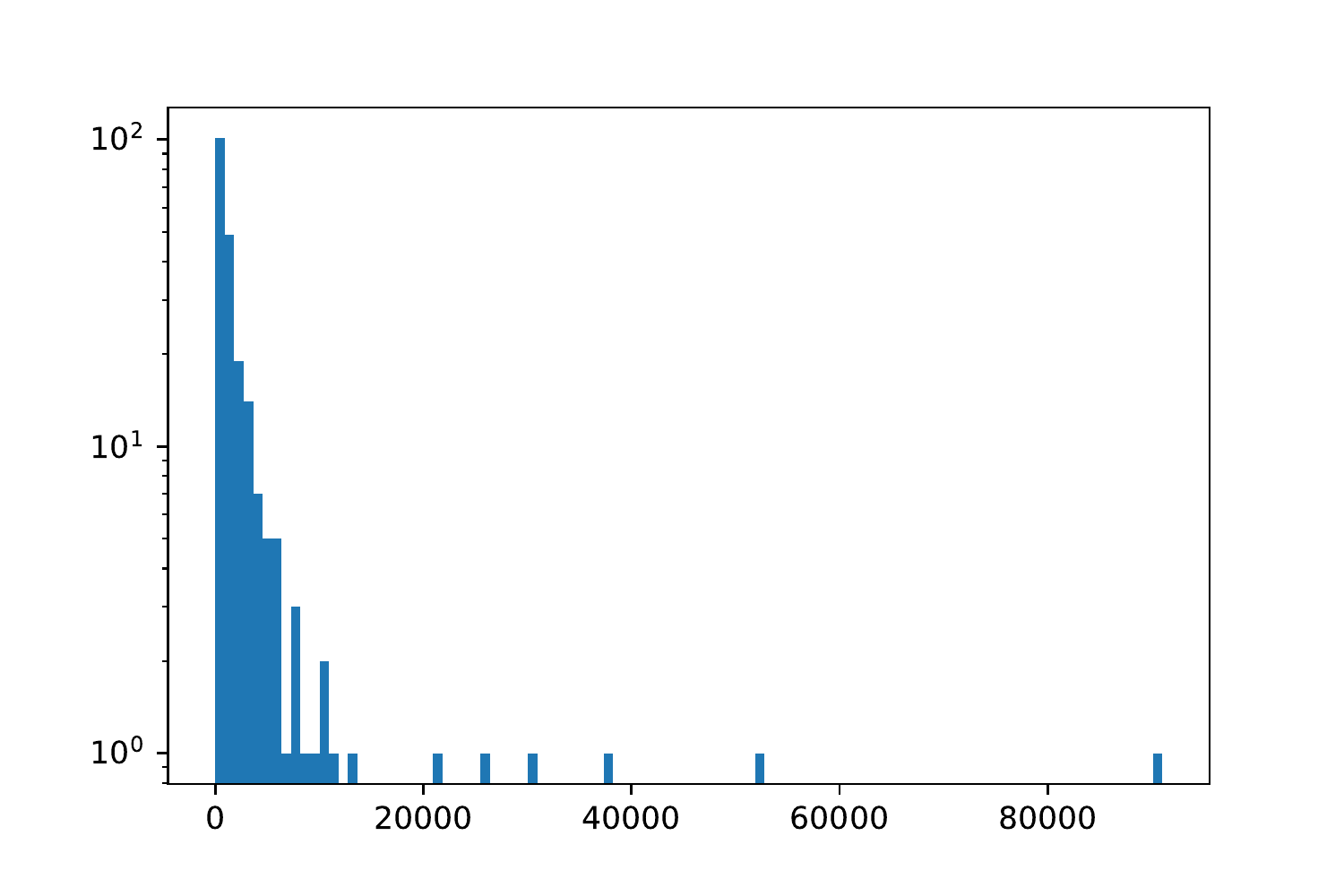}
  \label{subfigure: daydist_miss_size}
}
\subfloat[Data reuse counts]{%
  \includegraphics[width=0.24\linewidth, height=2.5cm]{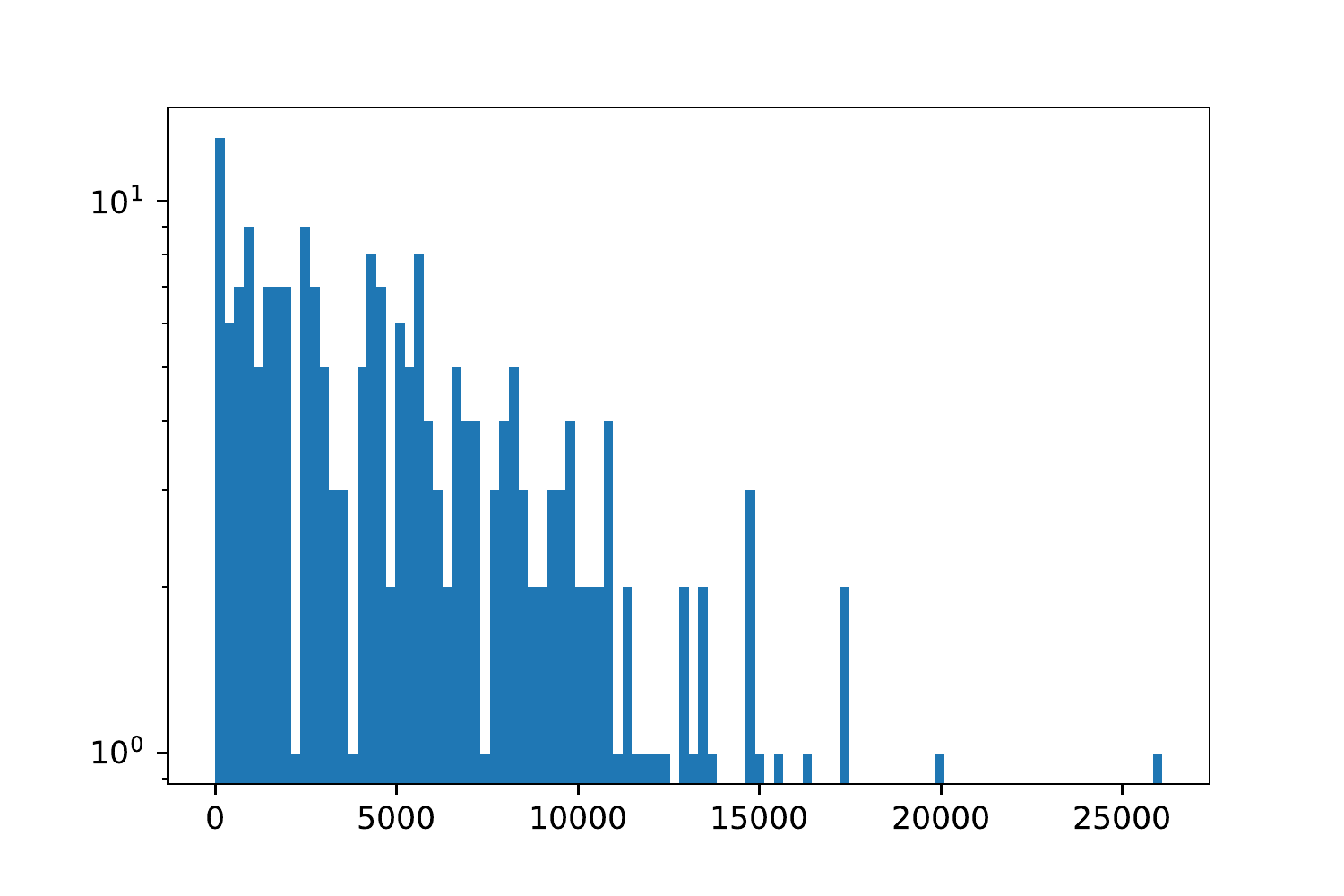}
  \label{subfigure: daydist_reuse_count}
}
\subfloat[Data reuse sizes]{%
  \includegraphics[width=0.24\linewidth, height=2.5cm]{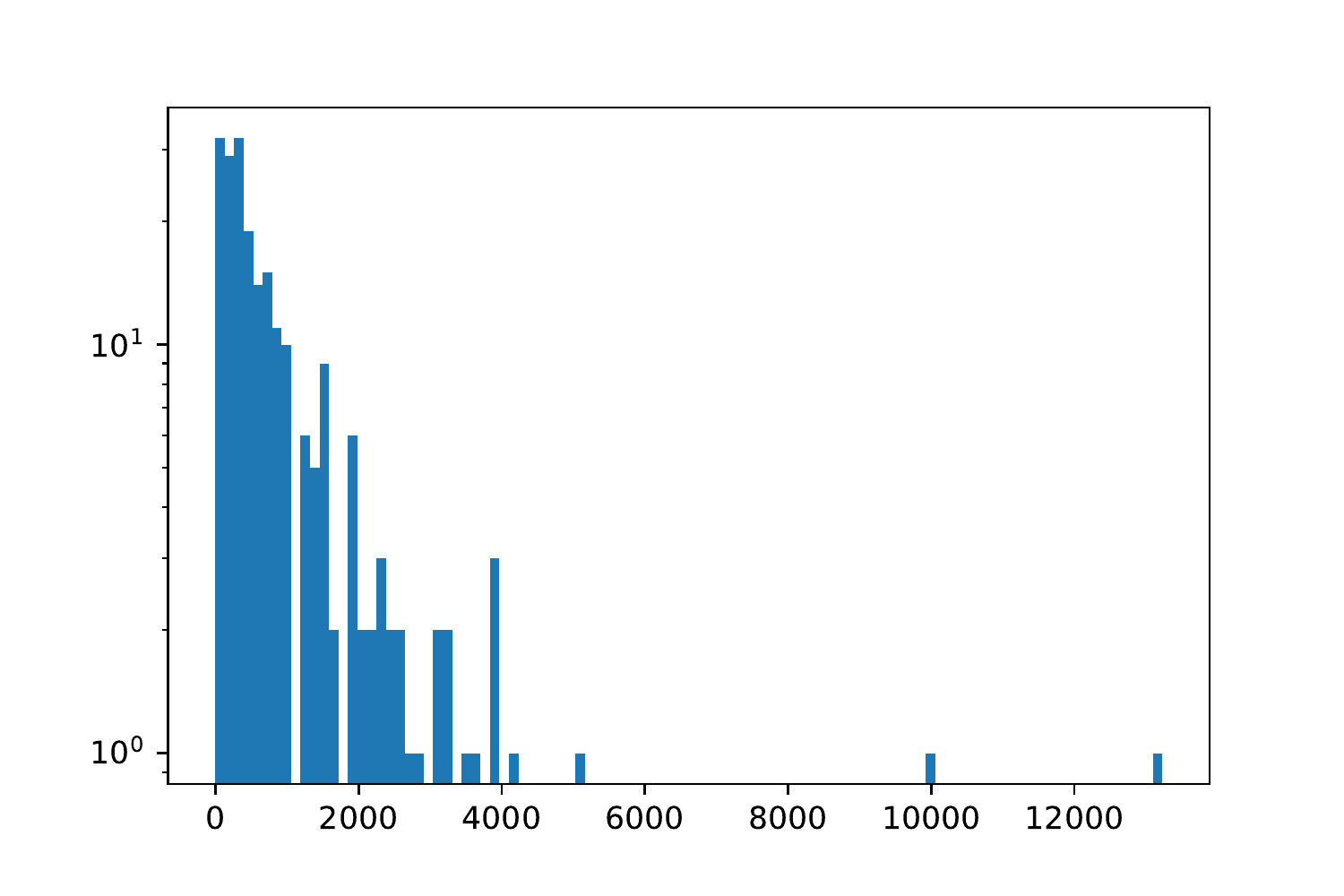}
  \label{subfigure: daydist_reuse_size}
} \newline
\caption{Distribution of daily features}
\label{fig:dayDist}
\end{figure*}

%%%%%%%%%%%%%%%%%%%%%%%%%%%%%%%%%%%%%%%%%%%%%%%%
\subsection{LSTM on the Daily Data}
We anticipate this modeling effort to be used in an advanced software-defined networking environment for possible resource allocation of a series of in-network caches.
In this context, one useful time frame for considering possible resource allocation might be a few hours or a day.
With this in mind, this work aggregates the cache utilization statistics into daily records.
To construct this daily time series, we need to generate meaningful daily summaries along with other useful features that might support the prediction task.
The daily summary of cache statistics includes the following features: (a) access counts, (b) access sizes, (c) cache hit counts, (d) cache hit sizes, (e) cache miss counts, (f) cache miss sizes, (g) data reuse counts, and (h) data reuse sizes.
%There exists a specific pattern in the data access data. This pattern can be partly extracted by recurrent neural network(RNN). The specifc RNN we use is the Long-Short Term Memory (LSTM). 3 models are built separately on the daily access data, hourly access data, and daily access data with 7-day moving average.

%\subsection{Data Preparation}
%Our LSTM models consist of the sum of each value of all cache nodes for the following features: (1) access counts, (2) access sizes, (3) cache hit counts, (4) cache hit sizes, (5) cache miss counts, (6) cache miss sizes, (7) data reuse counts, and (8) data reuse sizes.  

%Figure \ref{fig:hourDist} shows that the hourly data has a long-tail distribution. 
Figure \ref{fig:dayDist} shows the distribution of these daily summaries.
Since these features have widely varying values, we plan to normalize these values before giving them to LSTM models.
As there are many extreme values in the data, we have selected to use the z-score normalization~\cite{saranya2013study} instead of the more commonly used min-max normalization.

Due to the limited number of data points available, We allocate the data of the first 80\% of the study period to be the training data, and the data of the last 20\% of the study period to be the test data. The model selection would be based on how the model performs on the test data.
The train dataset covers from July 1, 2021 to Dec. 16, 2021, and the test dataset covers from Dec. 19, 2021 to Jan. 29, 2022.

%The train period is from July 1, 2021 to Dec. 16, 2021, and the test period is from Dec. 19, 2021 to Jan. 29, 2022.
We prepared two different models, one with the above mentioned eight features and the second one with one additional feature, day-of-the-week.
Because most workplaces follow the workweek schedule, we anticipate seeing a weekly trend and the day-of-the-week feature might improve the prediction accuracy. The day-of-the-week information is processed by one-hot encoding.    
%The day-of-the-week information is given from 0 to 6, representing from Monday to Sunday.

%As LSTM can capture the non-linear information in the data, the model can process the day-of-the-week information without using one-hot encoding.
%The day-of-the-week information is also normalized by z-score normalization.

The input of the daily LSTM model is a vector of size 8 or 14, depending on whether day-of-the-week information is added. The first 8 are the normalized features of $N_{th}$ day, and the features include data access count, data access size, cache hit count, cache hit size, cache miss count, cache miss size, data reuse count, data reuse sizes. The last 6 are used for one-hot encoding representation of the day-of-the-week information, indicating whether of $N_{th}$ day is Monday to Saturday. If $N_{th}$ day is Sunday, then it's represented as not Monday to Saturday.

The output of the LSTM model is a vector of size 8, the predicted normalized features of ${(N+1)}_{th}$ day, and the features include data access count, data access size, cache hit count, cache hit size, cache miss count, cache miss size, data reuse count, and data reuse sizes.
%The input of the daily LSTM model is a vector of size 9: the normalized data access count of $N_{th}$ day, the normalized data access size of $N_{th}$ day, the normalized cache hit count of $N_{th}$ day, the normalized cache hit size of $N_{th}$ day, the normalized cache miss count of $N_{th}$ day, the normalized cache miss size of $N_{th}$ day, the normalized data reuse count of $N_{th}$ day, the normalized data reuse sizes of $N_{th}$ day, and the normalized day-of-the-week of $N_{th}$ day. The output of the LSTM model is a vector of size 8: the normalized data access count of ${N+1}_{th}$ day, the normalized data access size of ${N+1}_{th}$ day, the normalized cache hit count of ${N+1}_{th}$ day, the normalized cache hit size of ${N+1}_{th}$ day, the normalized cache miss count of ${N+1}_{th}$ day, the normalized cache miss size of ${N+1}_{th}$ day, the normalized data reuse count of ${N+1}_{th}$ day, and the normalized data reuse sizes of ${N+1}_{th}$ day.
The loss function is the root mean squared error (RMSE).
All values in output vectors are given equal weights in calculating the loss.

\begin{table}%[h!]
\scriptsize
\centering
%\caption{Explored Hyper-parameters for Daily LSTM model}
\caption{Hyper-parameters for Daily LSTM model}
\begin{tabular}{|c||c|} \hline
parameter & values \tabularnewline \hline \hline
{ \# of first layer LSTM unit} & 16, 32, 64, 128, 256 \tabularnewline \hline
{ \# of second layer LSTM unit} & 0, 16, 32, 64, 128, 256 \tabularnewline \hline
first layer activation function & tanh, relu\tabularnewline \hline
second layer activation function & tanh, relu\tabularnewline \hline
dropout rate &  0, 0.04, 0.1, 0.15\tabularnewline \hline
{ \# of epochs}  & 5, 10, 15, 25, 50, 75,100 \tabularnewline \hline
\end{tabular}
\label{tab:daily_LSTM_comb}
\vspace{-0.2cm}
\end{table}

\begin{figure}%[H]
\centering 
\subfloat[]{%
  \includegraphics[width=0.49\linewidth]{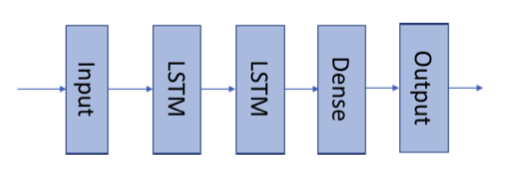}
  \label{subfigure:2layer_LSTM}
}
\subfloat[]{%
  \includegraphics[width=0.49\linewidth]{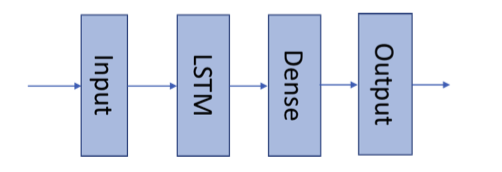}
  \label{subfigure:1layer_LSTM}
} \newline
\caption{(a) 2-layer LSTM (b) 1-layer LSTM }
\label{fig:LSTM_structure}
\end{figure}

Table \ref{tab:daily_LSTM_comb} shows the 3360 combinations of hyper-parameters explored for tuning the daily LSTM model. As we have a limited number of data points, the explored models have a maximum of 2 LSTM layers, and each LSTM layer has a maximum of 256 LSTM units. The structure of the daily LSTM is shown in Figure \ref{subfigure:2layer_LSTM}. When the number of the second layer LSTM unit is 0, the second LSTM layer does not exist; in this case, the daily LSTM is shown in Figure \ref{subfigure:1layer_LSTM}. 
The hyper-parameter of the final daily LSTM model is chosen by the RMSE %mean squared error 
between the predicted test set values and the true test set values. The final model with the lowest RMSE for the test set %mean squared error 
is a 1-layer LSTM model shown in Figure \ref{subfigure:1layer_LSTM}; its hyper-parameters are shown in table \ref{tab:daily_LSTM_para}.

\begin{table}%[h!]
\scriptsize
\centering
\caption{hyper-parameter of the daily LSTM model}
\begin{tabular}{|c||c|c|c|c|} \hline
  & {\# of LSTM unit} & {activation function} & {dropout rate} & {\# of epochs} \tabularnewline \hline \hline
values & 128 & tanh & 0.04 & 50\tabularnewline \hline
\end{tabular}
\label{tab:daily_LSTM_para}
\end{table}

%\vspace{-1.0cm}

\begin{figure*}%[htb!]
\centering 
\subfloat[Access counts]{%
  \includegraphics[width=0.49\linewidth]{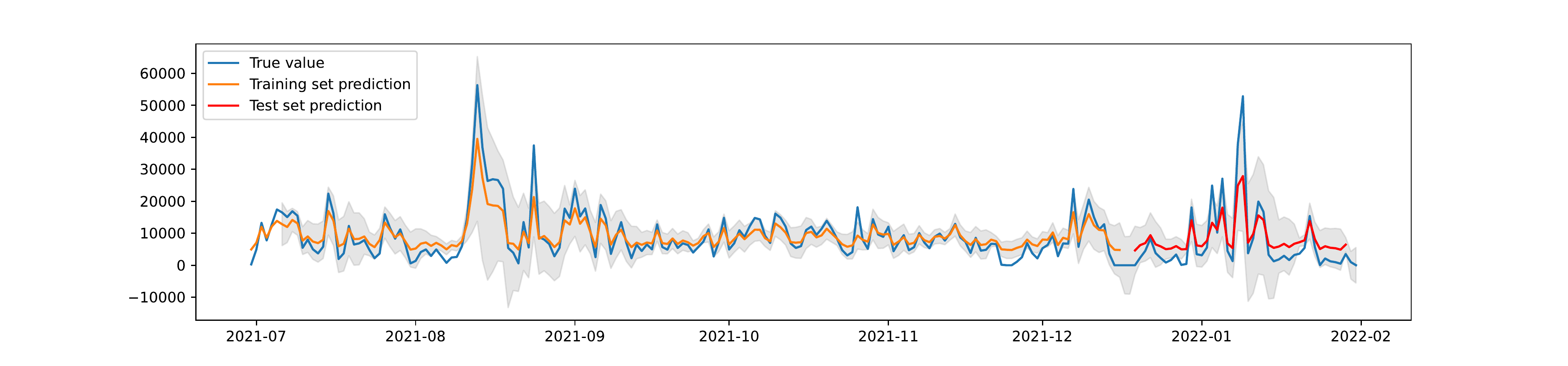}
  \label{subfigure:daily_access_count}
}
\subfloat[Access sizes]{%
  \includegraphics[width=0.49\linewidth]{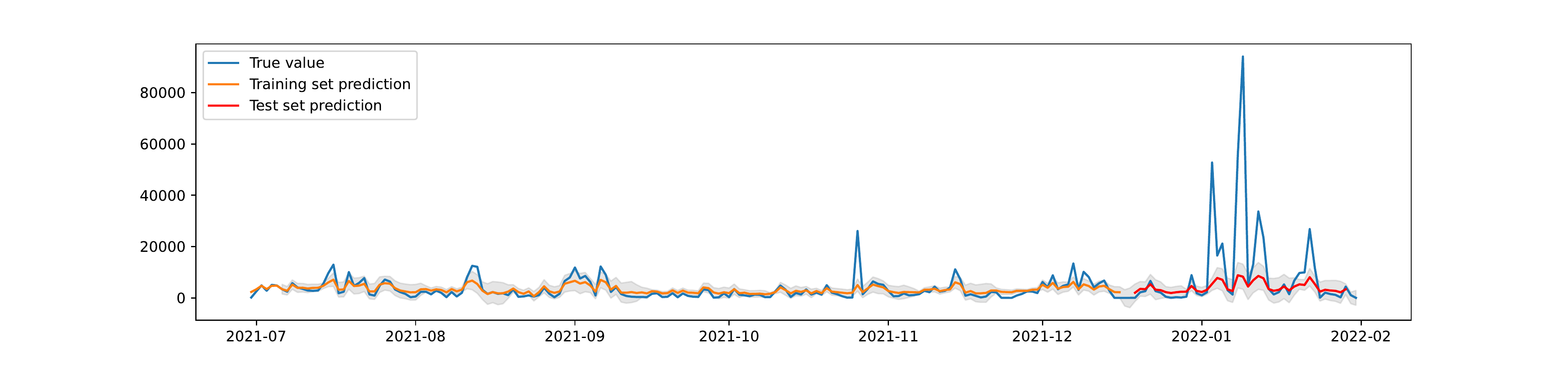}
  \label{subfigure:daily_access_size}
} \vspace{-0.5cm} \newline
\subfloat[Cache hit counts]{%
  \includegraphics[width=0.49\linewidth]{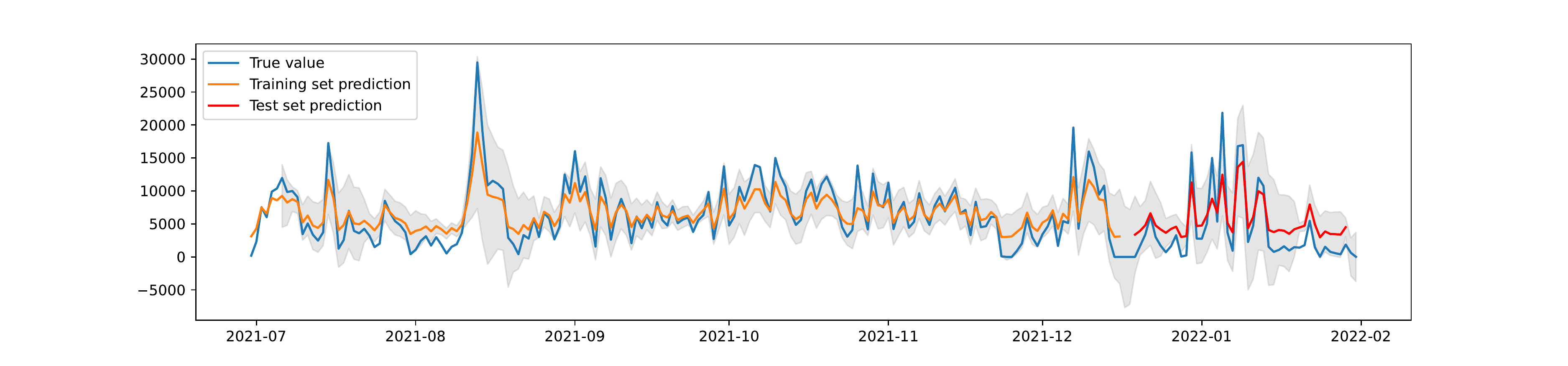}
  \label{subfigure:daily_hit_count}
}
\subfloat[Cache hit sizes]{%
  \includegraphics[width=0.49\linewidth]{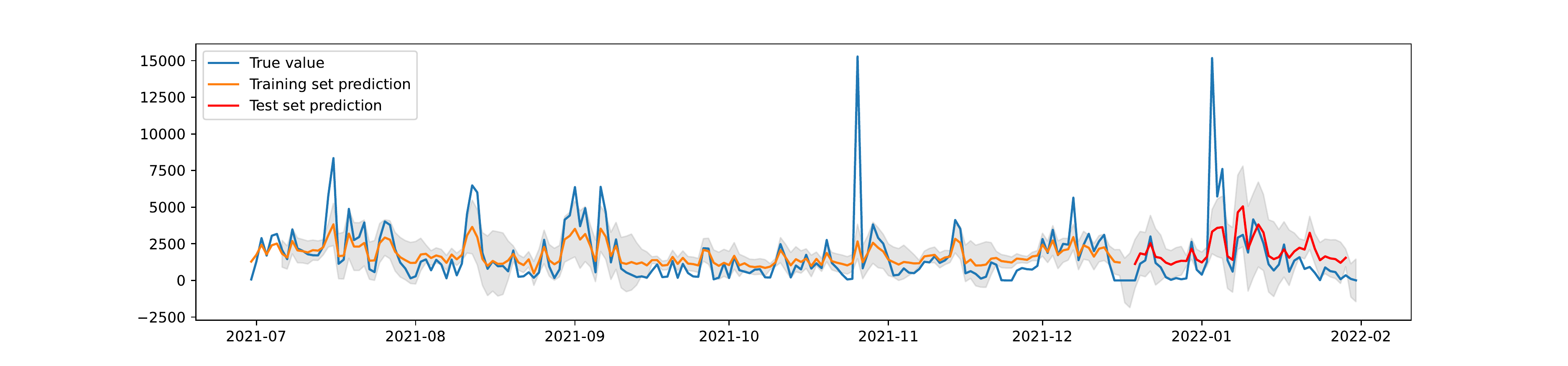}
  \label{subfigure:daily_hit_size}
} \vspace{-0.5cm} \newline
\subfloat[Cache miss counts]{%
  \includegraphics[width=0.49\linewidth]{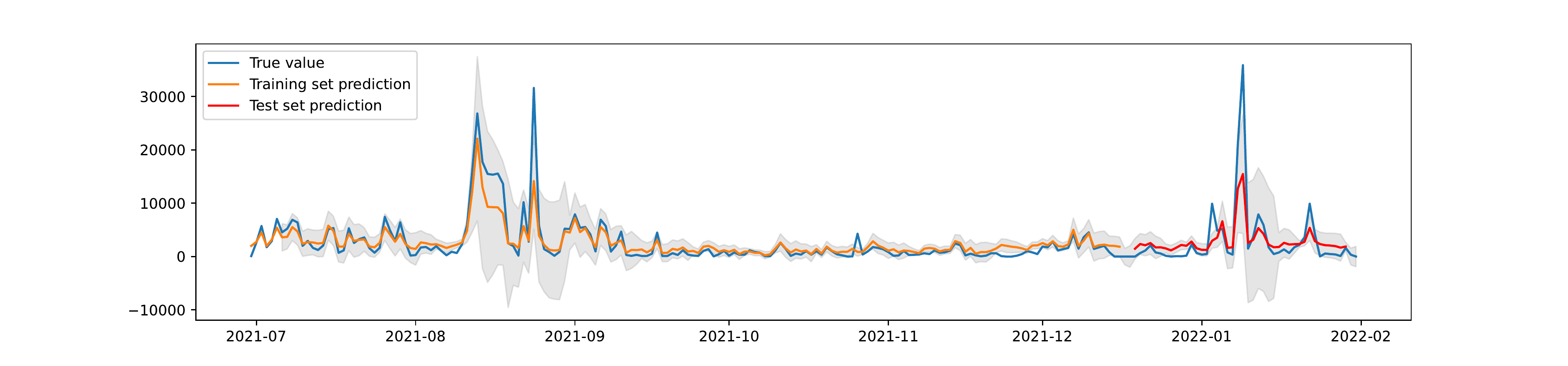}
  \label{subfigure:daily_miss_count}
}
\subfloat[Cache miss sizes]{%
  \includegraphics[width=0.49\linewidth]{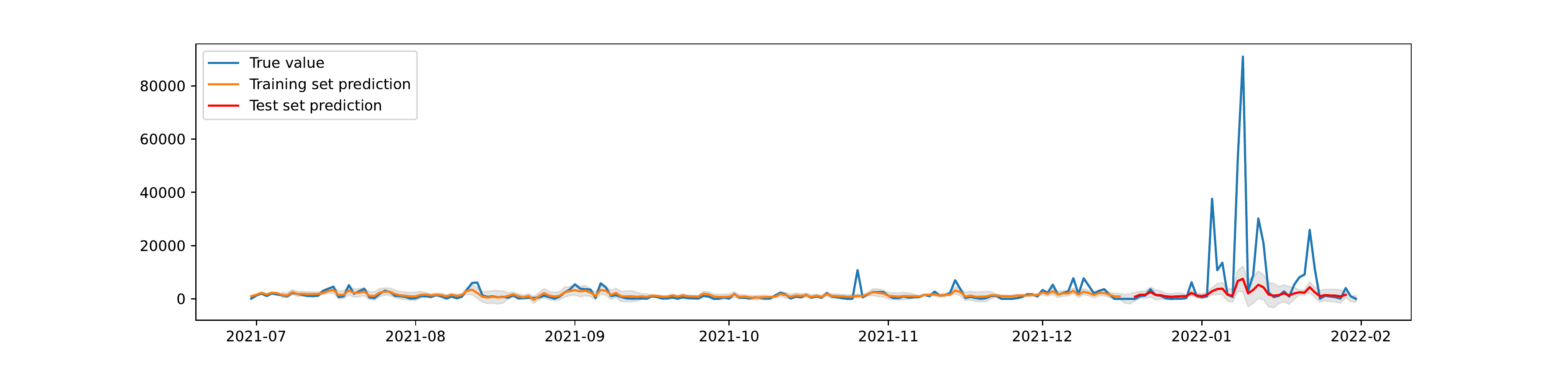}
  \label{subfigure:daily_miss_size}
} \vspace{-0.5cm} \newline
\subfloat[Data reuse counts]{%
  \includegraphics[width=0.49\linewidth]{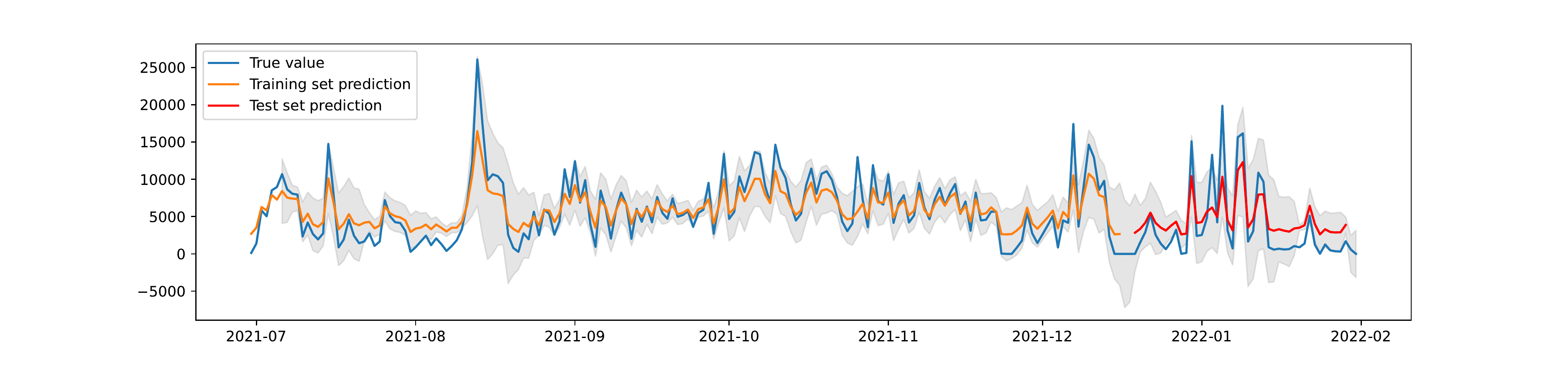}
  \label{subfigure:daily_reuse_count}
}
\subfloat[Data reuse sizes]{%
  \includegraphics[width=0.49\linewidth]{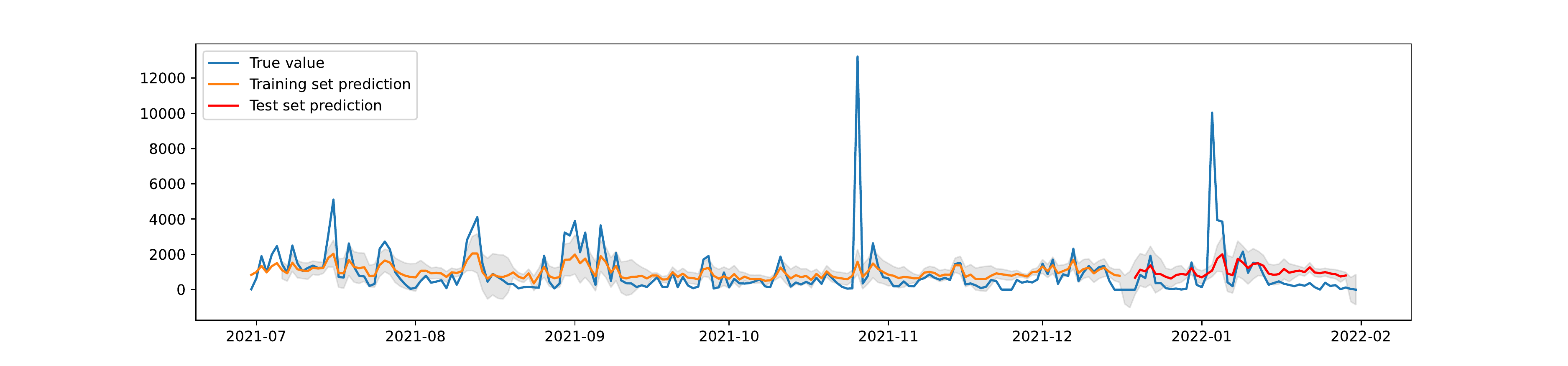}
  \label{subfigure:daily_reuse_size}
} \newline
\caption{Daily LSTM model Train and Test result vs True Value}
\label{fig:daily_LSTM_result}
\end{figure*}

Figure \ref{fig:daily_LSTM_result} shows how the daily LSTM model fits the daily access data. The model performs well when there are no extreme values, but as shown in Figure \ref{subfigure:daily_access_size}, \ref{subfigure:daily_hit_size}, \ref{subfigure:daily_miss_size}, and \ref{subfigure:daily_reuse_size}, the model does not fit and predict extreme values well. The gray shaded area is the predicted variance, defined as 2 standard deviations of the predicted values. If the actual value is within the predicted variance of the predicted value, we consider it as accurate. The overall accuracy is 0.884, and the accuracies for daily count data are all over 0.9.

\begin{table}%[h!]
\scriptsize
\centering
\caption{RMSE of Daily LSTM model with and without using weekday information}
\begin{tabular}{|c||c|c|c|c|c|} \hline
& \multicolumn{2}{|c|}{Without day-of-the-week} & \multicolumn{2}{|c|}{With day-of-the-week} & Acc. \tabularnewline \hline 
 & \shortstack{Train RMSE} & \shortstack{Test RMSE} & \shortstack{Train RMSE} & \shortstack{Test RMSE} \tabularnewline \hline \hline
Access Count & 3,861.14 & 4,944.34 & 3,492.61 & 4,220.19 & 0.93\tabularnewline \hline
Access Size & 2,480.61 & 16,621.57 & 2,612.90 & 16,571.21 & 0.85\tabularnewline \hline
Cache Hit Count & 2,459.72 & 3,158.99 & 2,179.03 & 2,917.99 & 0.95 \tabularnewline \hline
Cache Hit Size & 1,425.66 & 2,144.92 & 1,375.42 & 2,154.87 & 0.85  \tabularnewline \hline
Cache Miss Count & 2,261.62 & 2,954.13 & 2,302.29 & 2,970.10 & 0.91 \tabularnewline \hline
Cache Miss Size & 1,265.84 & 17,324.68 & 1,298.15 & 16,426.95 & 0.90 \tabularnewline \hline
Data Reuse Count & 2,224.82 & 3,066.91 & 2,063.65 & 2,646.69 & 0.93 \tabularnewline \hline
Data Reuse Size & 1,135.80 & 1,482.21 & 1,099.14 & 1,466.38 & 0.73 \tabularnewline \hline
\end{tabular}
\label{tab:daily_LSTM_rmse}
\vspace{-0.3cm}
\end{table}

Table \ref{tab:daily_LSTM_rmse} shows the RMSE of the Daily LSTM model on each daily data, along with the accuracy of the prediction.
Note that the RMSE shown in this table is measured on the scale of the original values, not the normalized values. %-- is this right? [JH]: Correct, RMSE in the table is calculated on the data that is converted back in its original scale.
The overall accuracy is 0.884.
The difference between the train RMSE and test RMSE on the size features is due to the model's inability to fit on extreme values.
When the day-of-the-week feature is added to the model for training, the model performance is improved on the daily counts, while the performance improvement in predicting daily sizes is minimal.
The extreme values in the daily sizes make it hard to fit the daily sizes well; thus, adding day-of-the-week information can only improve the performance on the daily counts.
This suggests that there might be a weekly seasonality in the daily data.

\subsection{LSTM on the Daily Data with 7-Day Moving Average (MA LSTM Model)}
In the previous study, we speculated that LSTM models perform poorly on the size feature because of the extreme values.
To verify this claim, we have smoothed the daily summaries with a 7-day moving average.
%on the daily data would greatly reduce the extreme values, and be able to eliminate the weekly seasonality in the daily data.%; thus, 7-day moving average on daily data may show a clear trend. The LSTM can perform nicely on the 7-day moving average on daily data.

%The train period is from 07/06/2021 to 12/18/2021, and the test period is from 12/21/2021 to 01/29/2022.

The input and output of the MA LSTM model are a vector of size 8, the normalized features of $N_{th}$ day and ${(N+1)}_{th}$ day respectively, and the features include data access count, data access size, cache hit count, cache hit size, cache miss count, cache miss size, data reuse count, and data reuse sizes.
%The input of the MA LSTM model is a vector of size 8:the normalized data access count of $N_{th}$ day, the normalized data access size of $N_{th}$ day, the normalized data shared count of $N_{th}$ day, the normalized data shared size of $N_{th}$ day, the normalized data transferred count of $N_{th}$ day, the normalized data transferred size of $N_{th}$ day, the normalized data reuse count of $N_{th}$ day, and the normalized data reuse sizes of $N_{th}$ day. The output of the LSTM model is a vector of size 8: the normalized data access count of ${N+1}_{th}$ day, the normalized data access size of ${N+1}_{th}$ day, the normalized data shared count of ${N+1}_{th}$ day, the normalized data shared size of ${N+1}_{th}$ day, the normalized data transferred count of ${N+1}_{th}$ day, the normalized data transferred size of ${N+1}_{th}$ day, the normalized data reuse count of ${N+1}_{th}$ day, and the normalized data reuse sizes of ${N+1}_{th}$ day.
The loss function is the root mean squared error (RMSE). All values in the output vectors are given equal weights in calculating the loss.

The same 3360 combinations of hyper-parameters shown in Table \ref{tab:daily_LSTM_comb} are explored in the MA LSTM model. The model selection process is the same as the selection process for the daily LSTM model. The model with the lowest test RMSE is the 1-layer LSTM model shown in Figure \ref{subfigure:1layer_LSTM}; its hyper-parameters are shown in Table \ref{tab:MA_LSTM_para}. The hyper-parameters and constructions of the daily LSTM model and the MA LSTM model are very similar as they only differ in the dropout rate and the number of training epochs. This is due to the high similarity between the daily data and the daily data with 7-day moving average, and the limited number of available data points.

\begin{table}%[h!]
\scriptsize
\centering
\caption{hyper-parameter of the MA LSTM model}
\begin{tabular}{|c||c|c|c|c|} \hline
  & {\# of LSTM unit} & {activation function} & {dropout rate} & {\# of epochs} \tabularnewline \hline \hline
values & 128 & tanh & 0.00 & 100 \tabularnewline \hline
\end{tabular}
\label{tab:MA_LSTM_para}
\end{table}

\begin{figure*}%[htb!]
\centering 
\subfloat[Access counts]{%
  \includegraphics[width=0.49\linewidth, height=2.1cm]{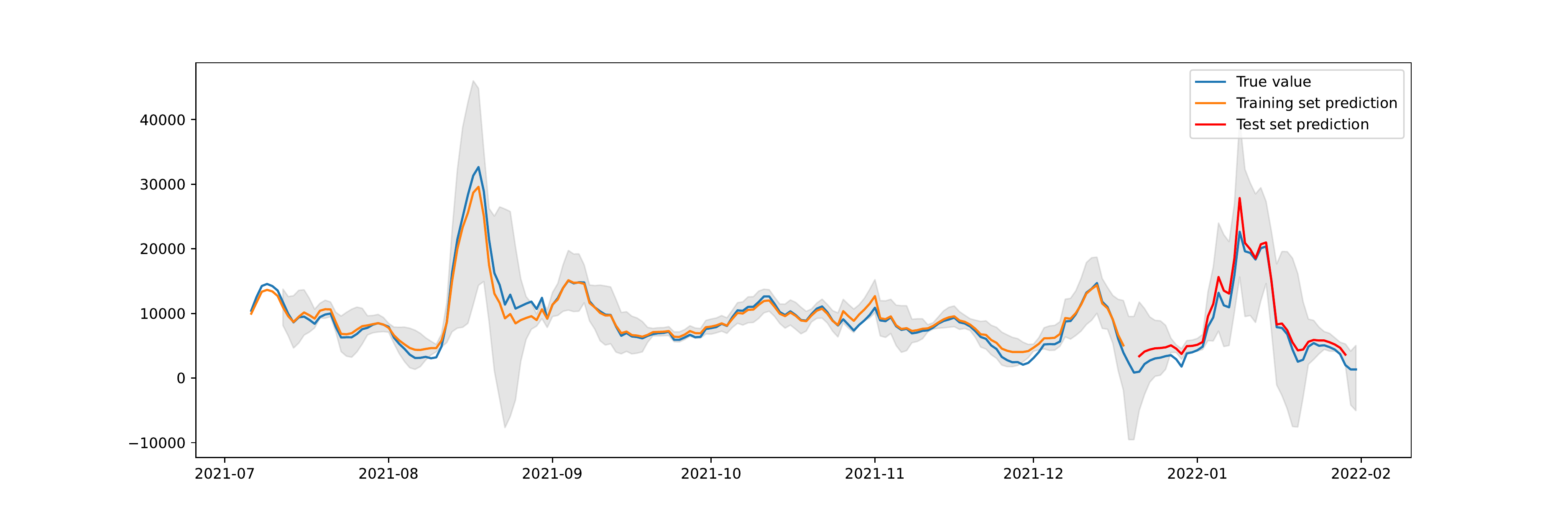}
  \label{subfigure:MA_access_count}
}
\subfloat[Access sizes]{%
  \includegraphics[width=0.49\linewidth, height=2.1cm]{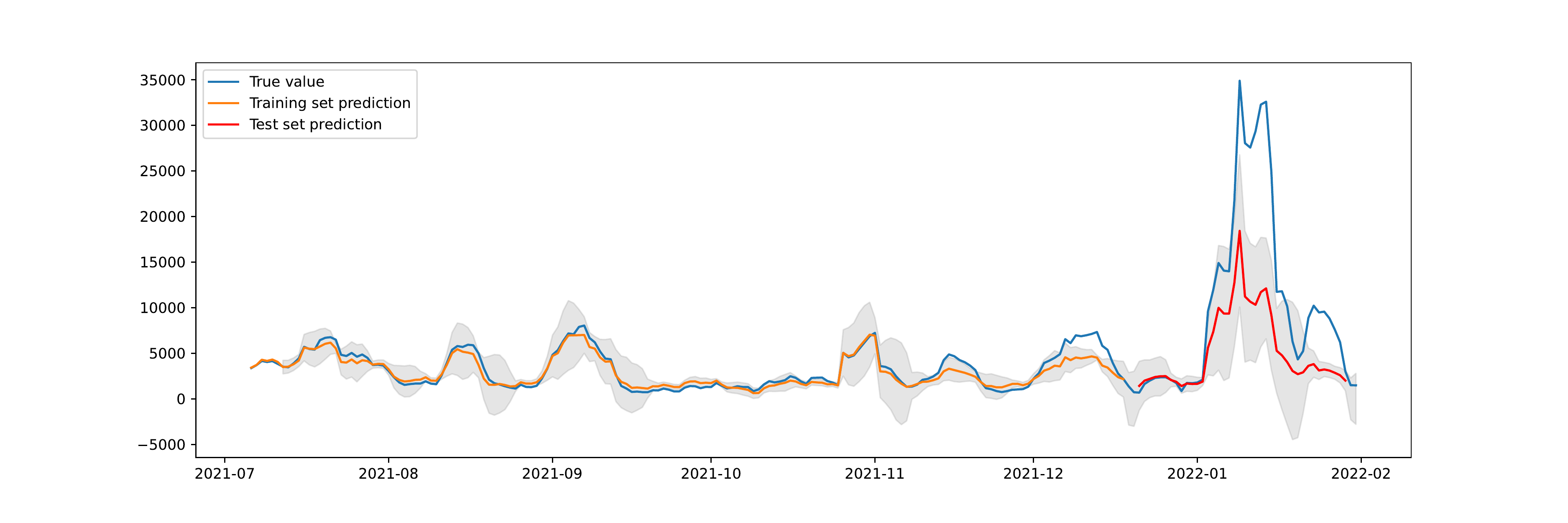}
  \label{subfigure:MA_access_size}
} \vspace{-0.5cm} \newline 
\subfloat[Cache hit counts]{%
  \includegraphics[width=0.49\linewidth, height=2.1cm]{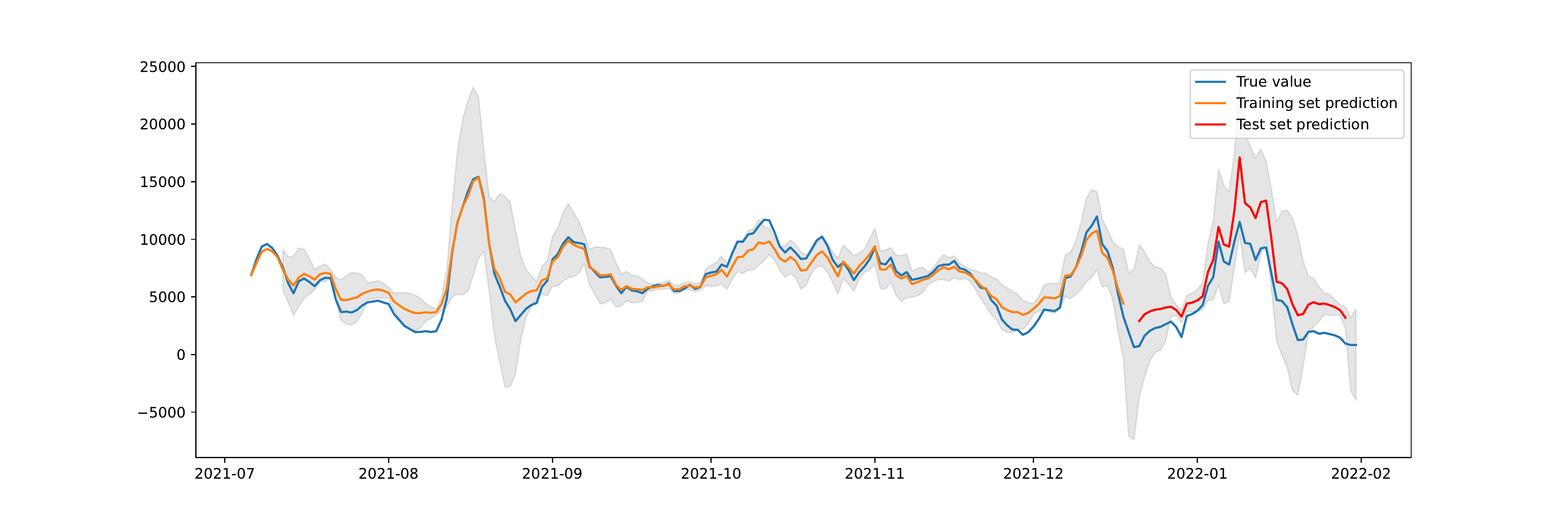}
  \label{subfigure:MA_hit_count}
}
\subfloat[Cache hit sizes]{%
  \includegraphics[width=0.49\linewidth, height=2.1cm]{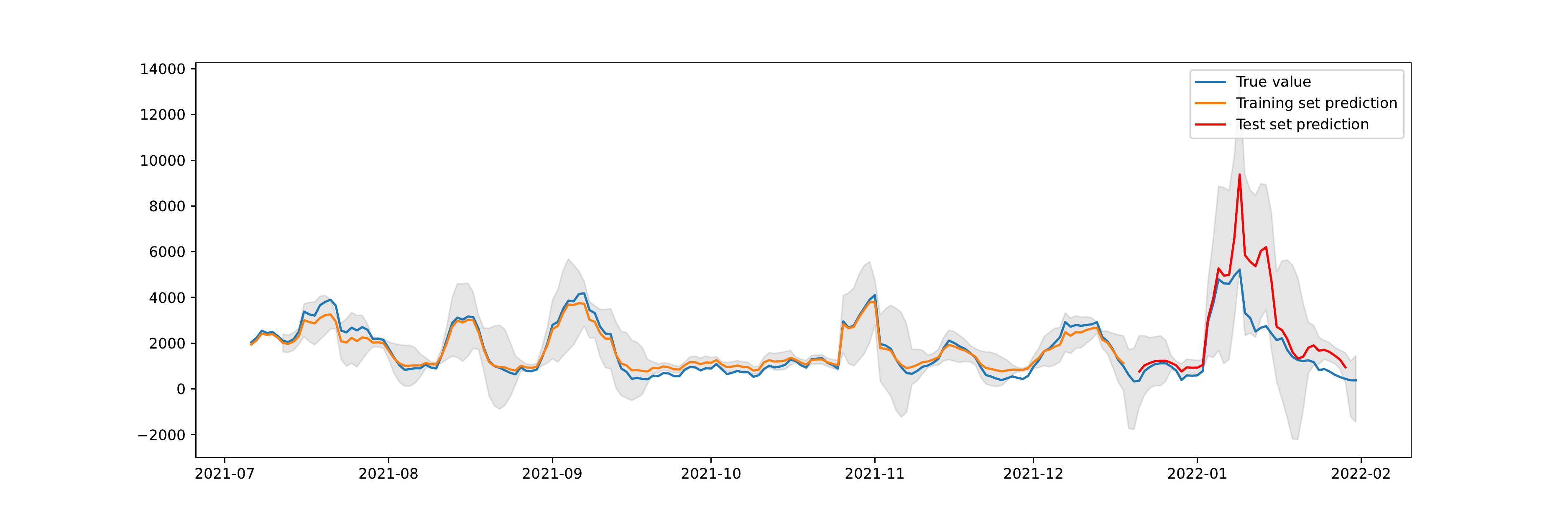}
  \label{subfigure:MA_hit_size}
} \vspace{-0.5cm} \newline
\subfloat[Cache miss counts]{%
  \includegraphics[width=0.49\linewidth, height=2.1cm]{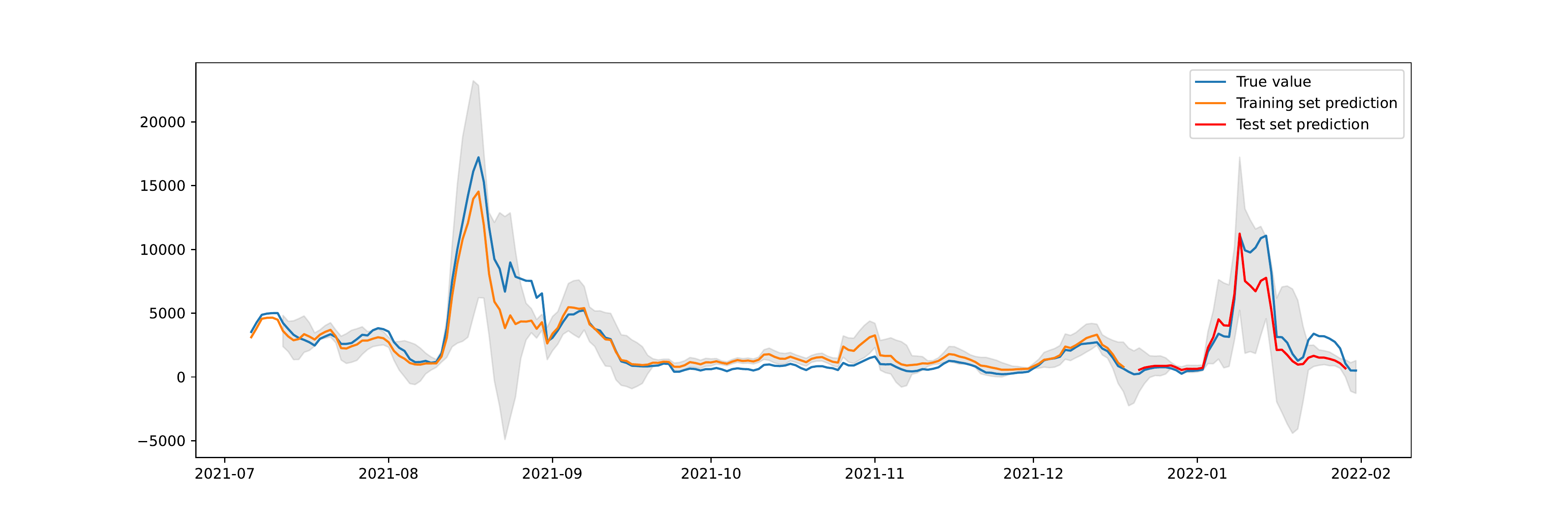}
  \label{subfigure:MA_miss_count}
}
\subfloat[Cache miss sizes]{%
  \includegraphics[width=0.49\linewidth, height=2.1cm]{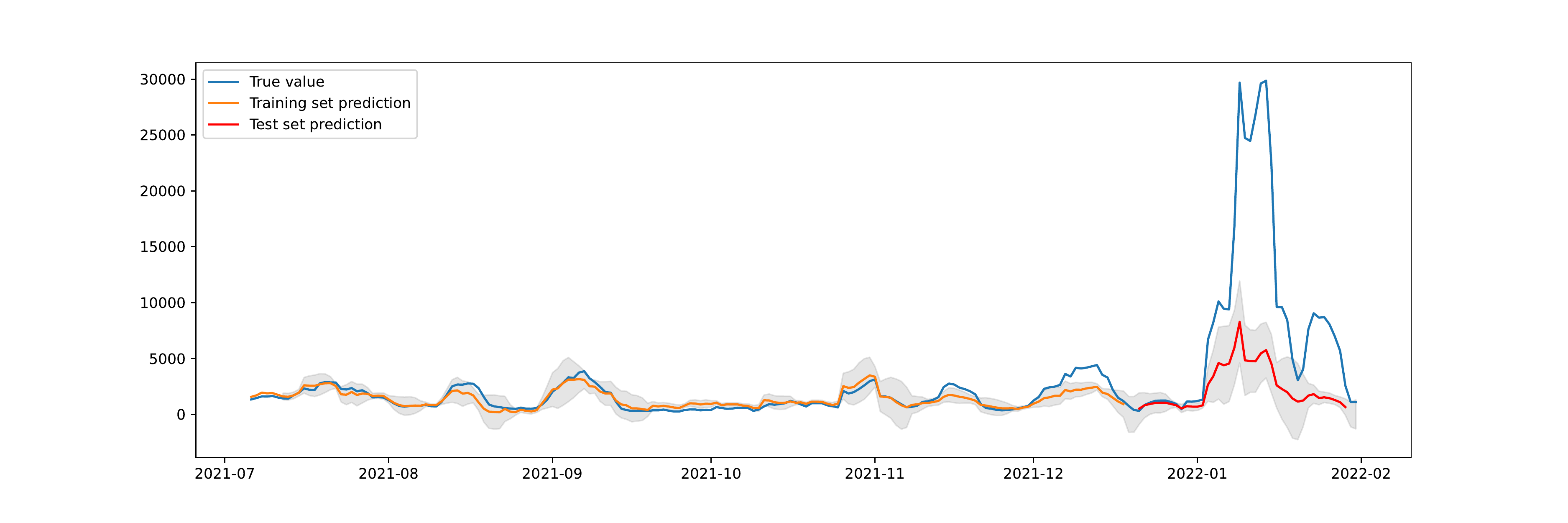}
  \label{subfigure:MA_miss_size}
} \vspace{-0.5cm} \newline
\subfloat[Data reuse counts]{%
  \includegraphics[width=0.49\linewidth, height=2.1cm]{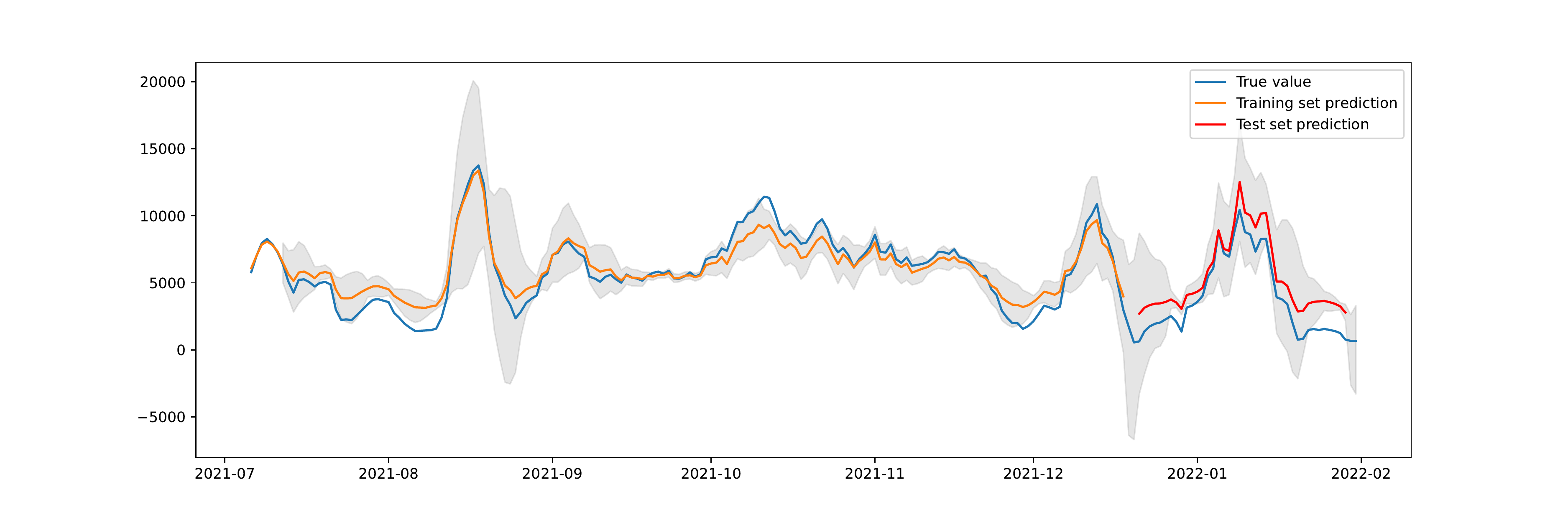}
  \label{subfigure:MA_reuse_count}
}
\subfloat[Data reuse sizes]{%
  \includegraphics[width=0.49\linewidth, height=2.1cm]{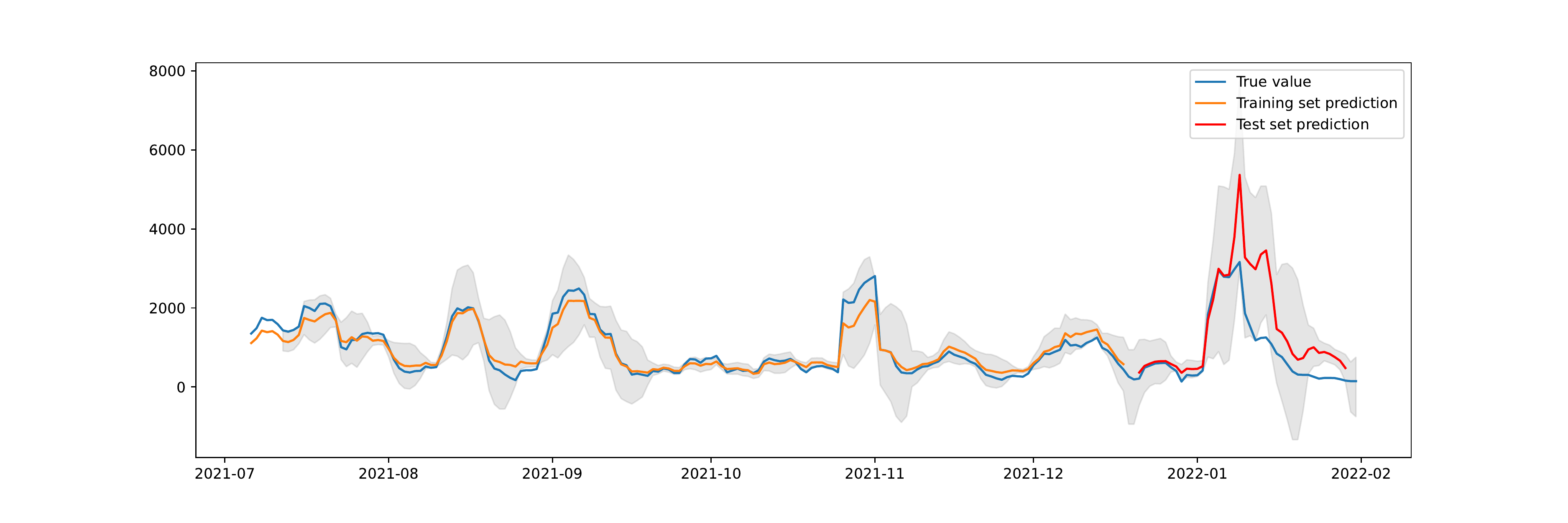}
  \label{subfigure:MA_reuse_size}
} \newline
\caption{MA LSTM model Train and Test result vs True Value}
\label{fig:MA_LSTM_result}
\end{figure*}

%\fix{[AS] The hyper-parameters are different in only in the number of units. Can we say something about the diff in the model?FIXED: In term of the model, I don't think there is to much differance. I will explain bit why this two models might be similar}
Figure \ref{fig:MA_LSTM_result} shows how the MA LSTM model fits the 7-day moving average on daily data. The model still deviates a lot on the extreme values in Figure \ref{subfigure:MA_miss_size}, but the model works well in general. The gray shaded area indicates the predicted variance, which is much smaller compared to the daily LSTM model.

\begin{table}%[h!]
\scriptsize
\centering
\caption{Explored Hyper-parameters for MA LSTM model}
\begin{tabular}{|c||c|c|c|c|} \hline
 & Train RMSE & Test RMSE & Accuracy & \shortstack{Test RMSE reduction \\ compare with daily LSTM} \tabularnewline \hline \hline
Access Count & 1,122.15 & 2,169.72 & 0.93 & 48.6\%\tabularnewline \hline
Access Size &  744.56 & 7,729.04 & 0.83 & 53.4\%\tabularnewline \hline
Cache Hit Count & 829.23 & 2025.21 & 0.88 & 30.6\%\tabularnewline \hline
Cache Hit Size &  223.00 & 1,573.72 & 0.91 & 27.1\%\tabularnewline \hline
Cache Miss Count & 1,127.30 & 781.83 & 0.86 & 73.0\%\tabularnewline \hline
Cache Miss Size & 612.94 & 9616.83 & 0.77 & 58.5\%\tabularnewline \hline
Data Reuse Count & 808.80 & 1,228.71 & 0.87 & 53.6\%\tabularnewline \hline
Data Reuse Size & 208.27 & 812.33 & 0.92 & 44.6\%\tabularnewline \hline
\end{tabular}
\label{tab:MA_LSTM_rmse}
\end{table}

Table \ref{tab:MA_LSTM_rmse} shows the RMSE of the MA LSTM model, along with the prediction accuracy. Overall accuracy is 0.873. Although accuracy is less than 0.01 lower than the daily LSTM model, the predicted variance of the MA LSTM model is much smaller, so the prediction of the MA LSTM model is closer to the actual value.
%\fix{I thought the overall accuracy would be better than that of Table 4? [JH]: The accuracy is calculated by the proportion of true values falls into the grey bands in the plots which are based on std of predicted values. Since the MA Model fluctuate not as much, the margin of the bands are much thinner, so the true values are less likely to fall in to the gray bands}
Compared to the RMSE of the daily LSTM model, the MA LSTM model performs much better overall in terms of the test set RMSE; 

%48.6\% reduction in access count, 53.4\% reduction in access size, 30.6\% reduction in cache hit count, 27.1\% reduction in cache hit size, 73\% reduction in cache miss count, 58.5\% reduction in cache miss size, 53.6\% reduction in data reuse count, and 44.6\% reduction in data reuse size.

This shows that the LSTM model fits the daily data with 7-day moving average better than the daily data, which confirms that the extreme values severely affect the LSTM performance.
% accuracy is within 3 STDs. Overall, 87.3\% of the true values fall into the 3 STDs of the predicted values.
%\fix{[AS] Some words on the comparing daily LSTM RMSE table and daily MA LSTM RMSE table. FIXED}

\iffalse
% backup
\begin{table}[h!]
\scriptsize
\centering
\caption{Explored Hyper-parameters for MA LSTM model}
\begin{tabular}{|c||c|c|} \hline
 & Train RMSE & Test RMSE \tabularnewline \hline \hline
Access Count & 1,122.15 & 2,169.72\tabularnewline \hline
Access Size &  744.56 & 7,729.04\tabularnewline \hline
Cache Hit Count & 829.23 & 2025.21 \tabularnewline \hline
Cache Hit Size &  223.00 & 1,573.72 \tabularnewline \hline
Cache Miss Count & 1,127.30 & 781.83 \tabularnewline \hline
Cache Miss Size & 612.94 & 9616.83\tabularnewline \hline
Data Reuse Count & 808.80 & 1,228.71 \tabularnewline \hline
Data Reuse Size & 208.27 & 812.33 \tabularnewline \hline
\end{tabular}
\label{tab:MA_LSTM_rmse}
\end{table}
\fi

%%%%%%%%%%%%%%%%%%%%%%%%%%%%%%%%%%%%%%%%%%%%%%%%
\subsection{Seasonality}
Day-of-the-week information improves the performance of the daily the LSTM model, which suggests some weekly seasonality in the daily time series data.
We investigate the seasonality using periodograms~\cite{periodogram_cite}.
%, we explored the period of seasonal effect in the data.
%As any time series can be constructed by a combination of cosine and sine waves with different periods and amplitudes. Periodogram can be used to identify the dominant periods of a time series. \cite{periodogram_cite}. 

%As adding day-of-the-week information and hour-of-the-day information can improve the performance of daily LSTM model and hourly LSTM model, there exist seasonality in the daily and hourly data. The use of Fast-Fourier Transformation can explore the period of seasonal effect in the data.

\begin{figure*}%[htb!]
\centering 
\subfloat[Access counts]{%
  \includegraphics[width=0.24\linewidth, height=2.5cm]{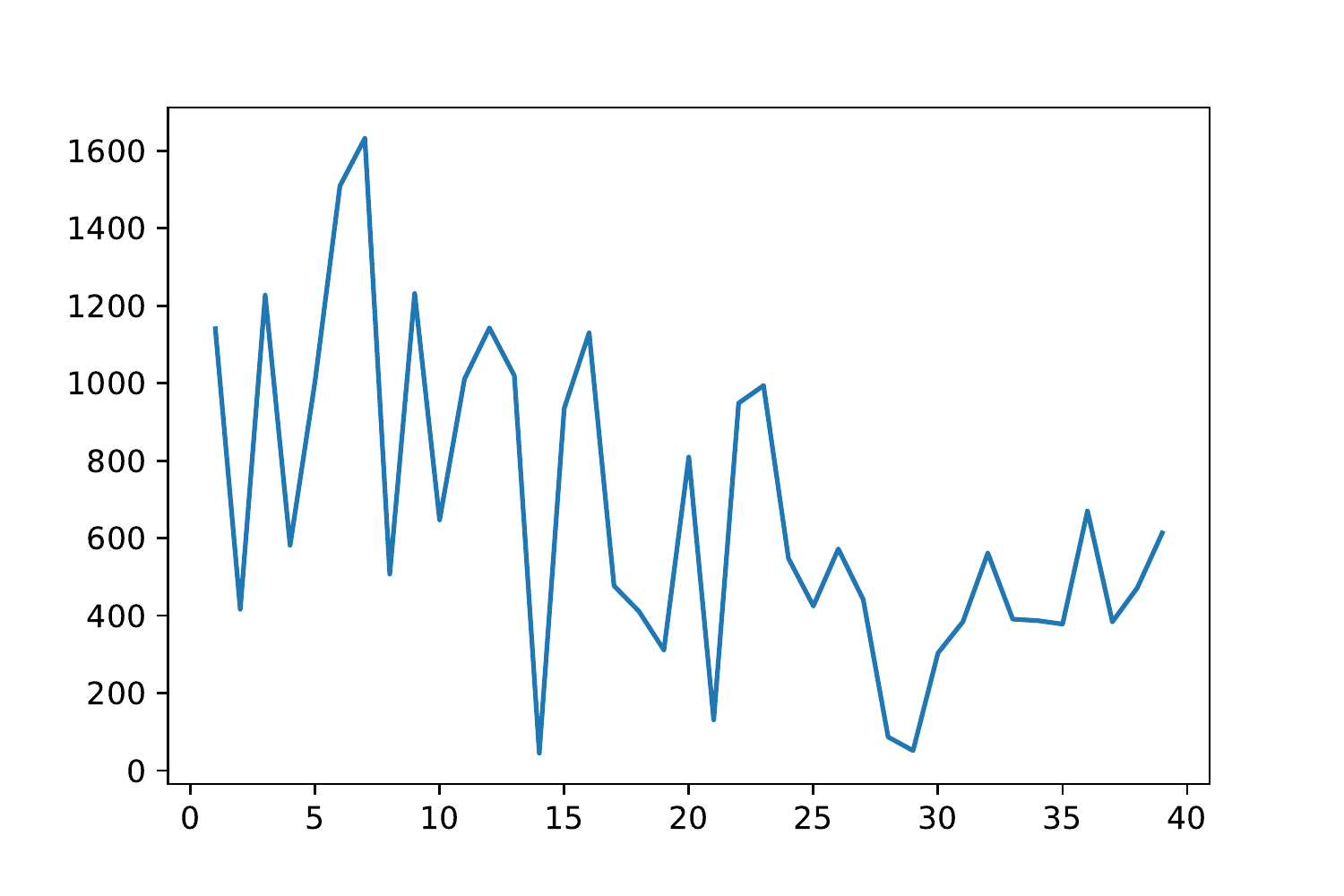}
  \label{subfigure: dayDFT_access_count}
}
\subfloat[Access sizes]{%
  \includegraphics[width=0.24\linewidth, height=2.5cm]{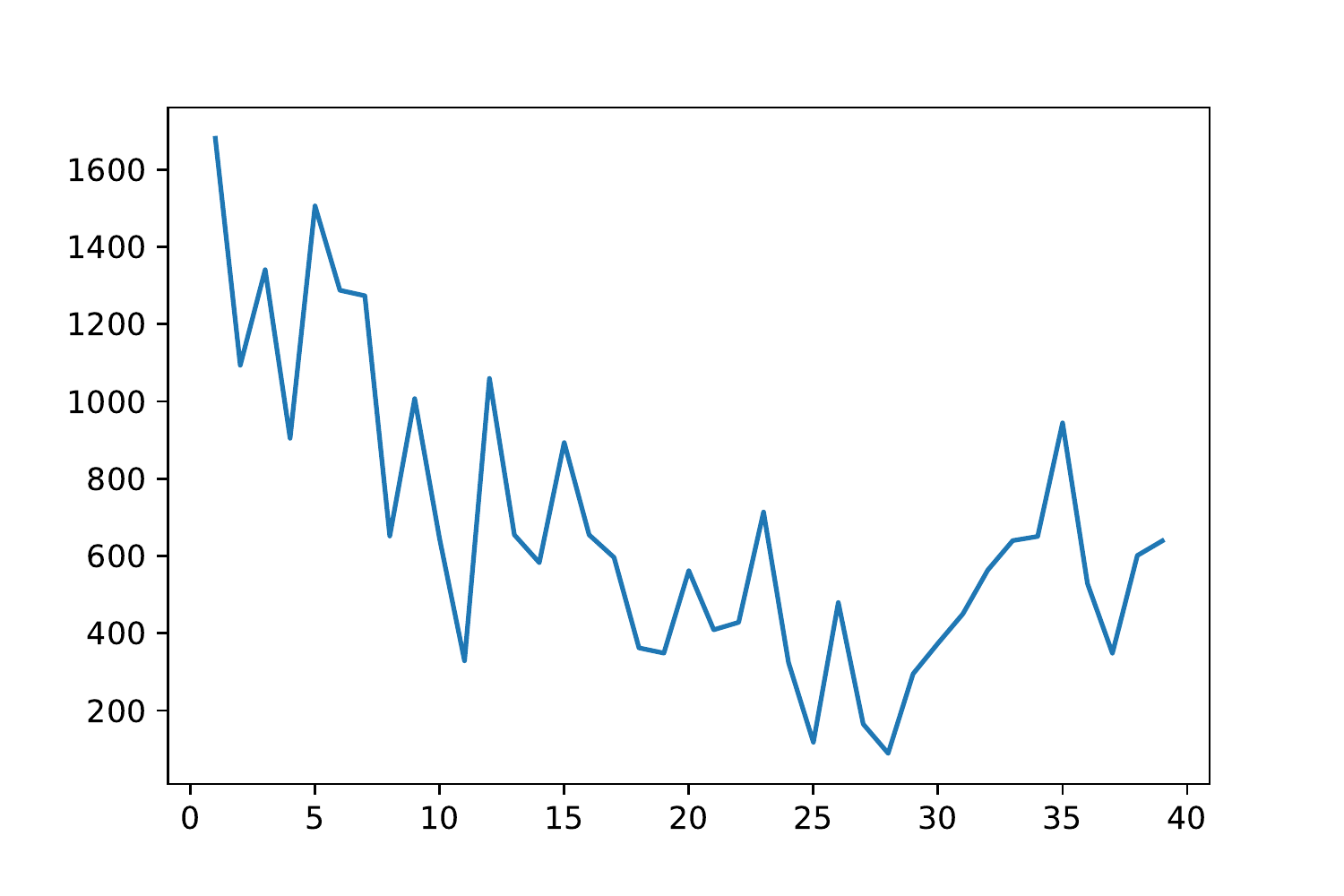}
  \label{subfigure: dayDFT_access_size}
}
\subfloat[Cache hit counts]{%
  \includegraphics[width=0.24\linewidth, height=2.5cm]{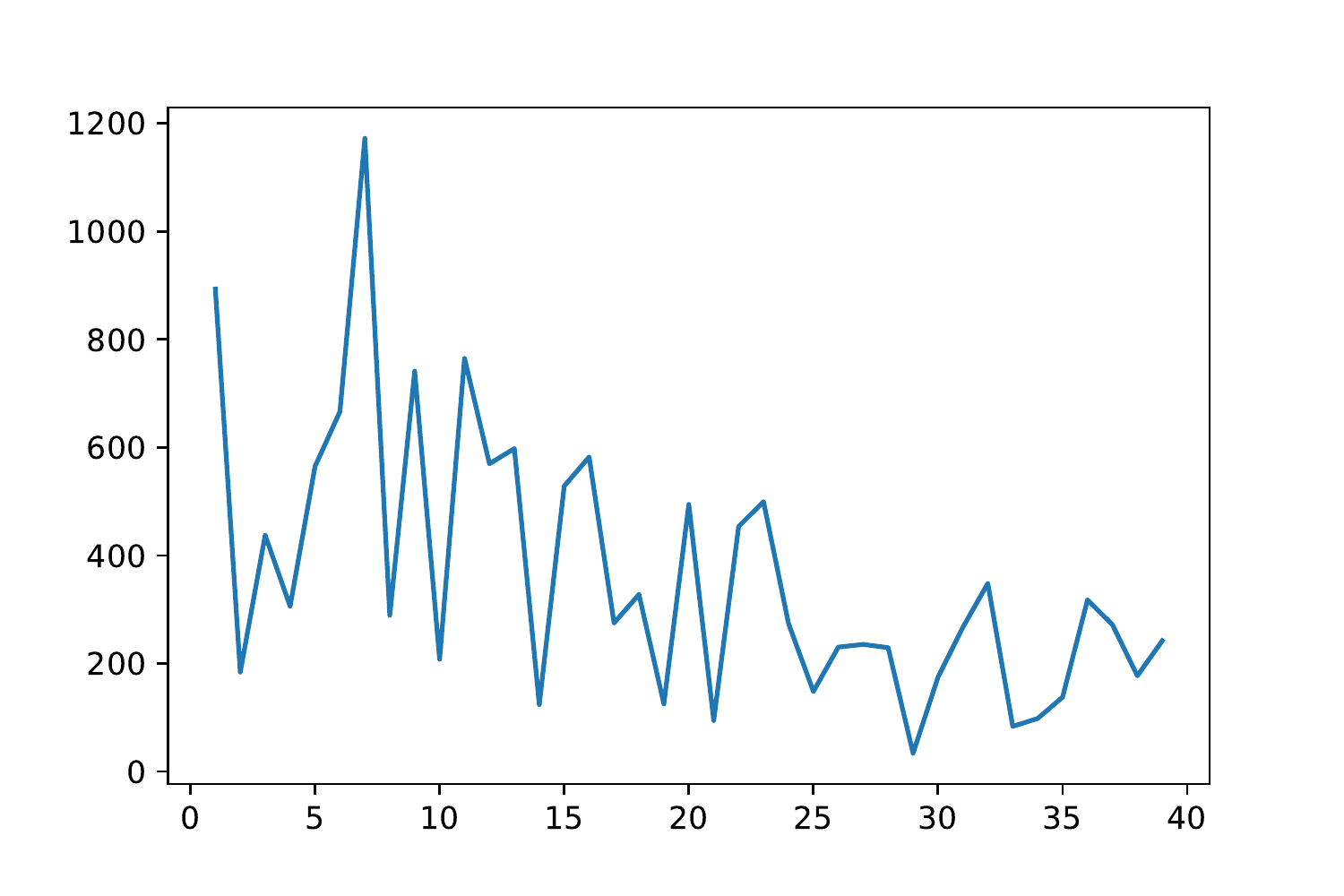}
  \label{subfigure: dayDFT_hit_count}
}
\subfloat[Cache hit sizes]{%
  \includegraphics[width=0.24\linewidth, height=2.5cm]{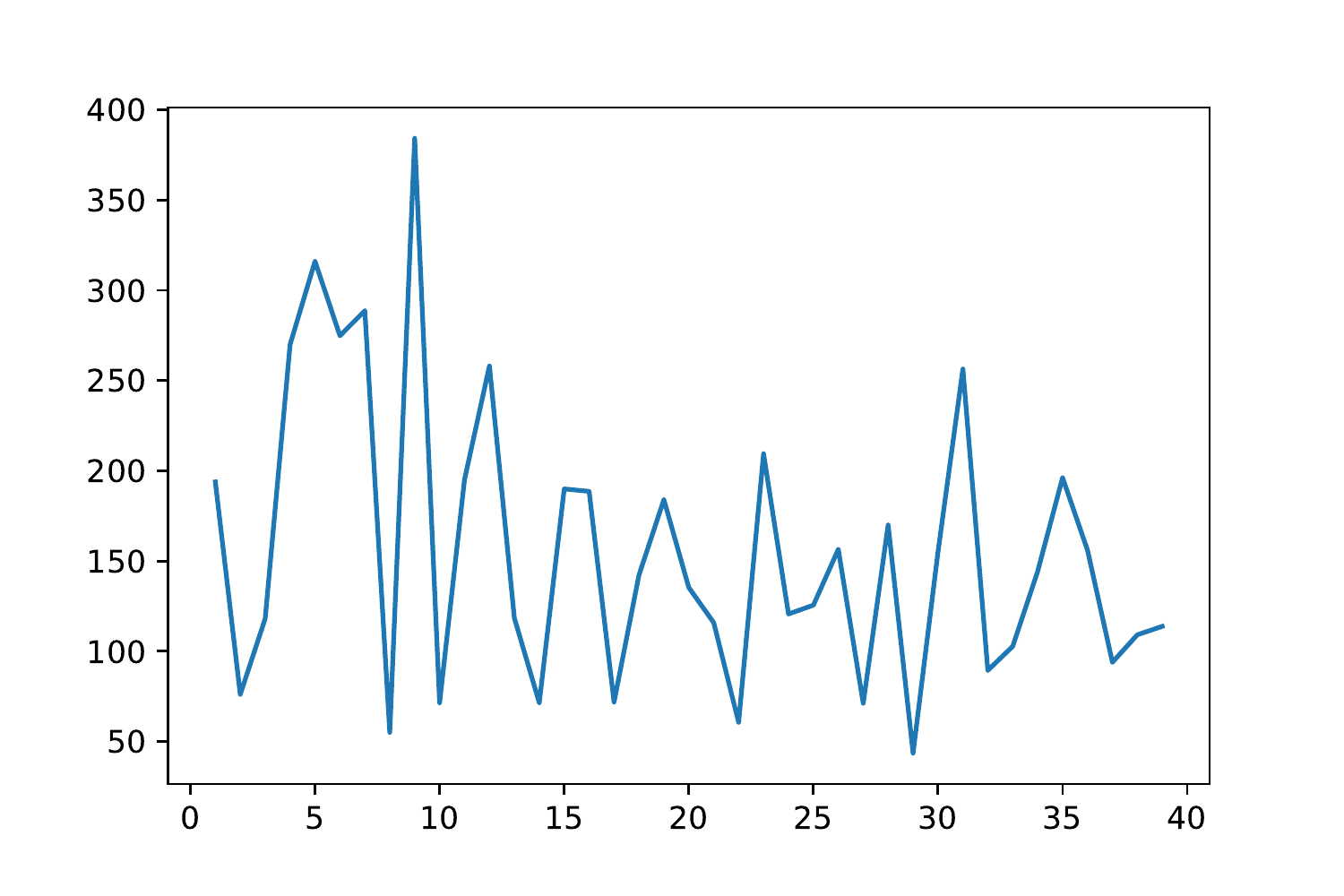}
  \label{subfigure: dayDFT_hit_size}
} \vspace{-0.5cm} \newline
\subfloat[Cache miss counts]{%
  \includegraphics[width=0.24\linewidth, height=2.5cm]{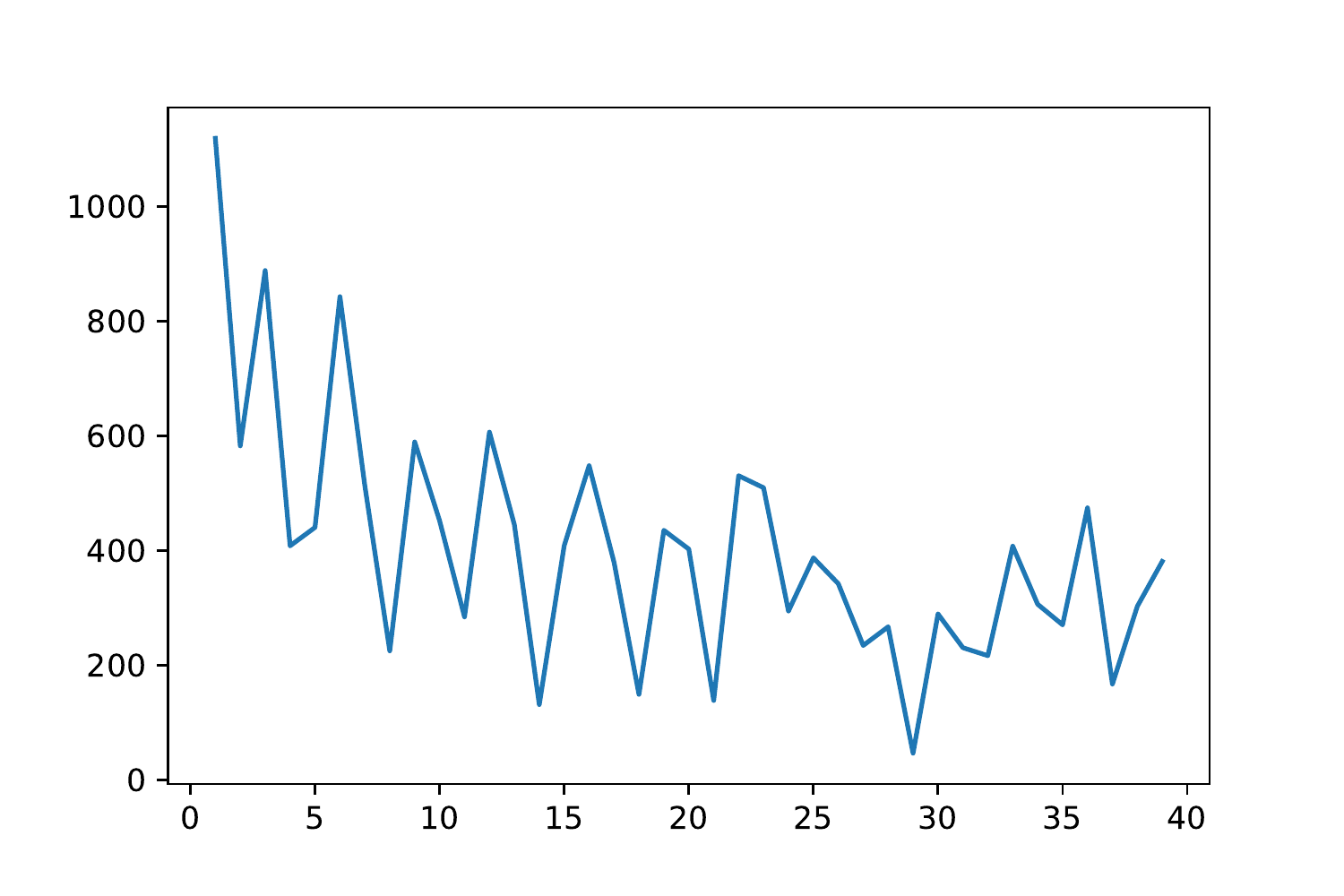}
  \label{subfigure: dayDFT_miss_count}
}
\subfloat[Cache miss sizes]{%
  \includegraphics[width=0.24\linewidth, height=2.5cm]{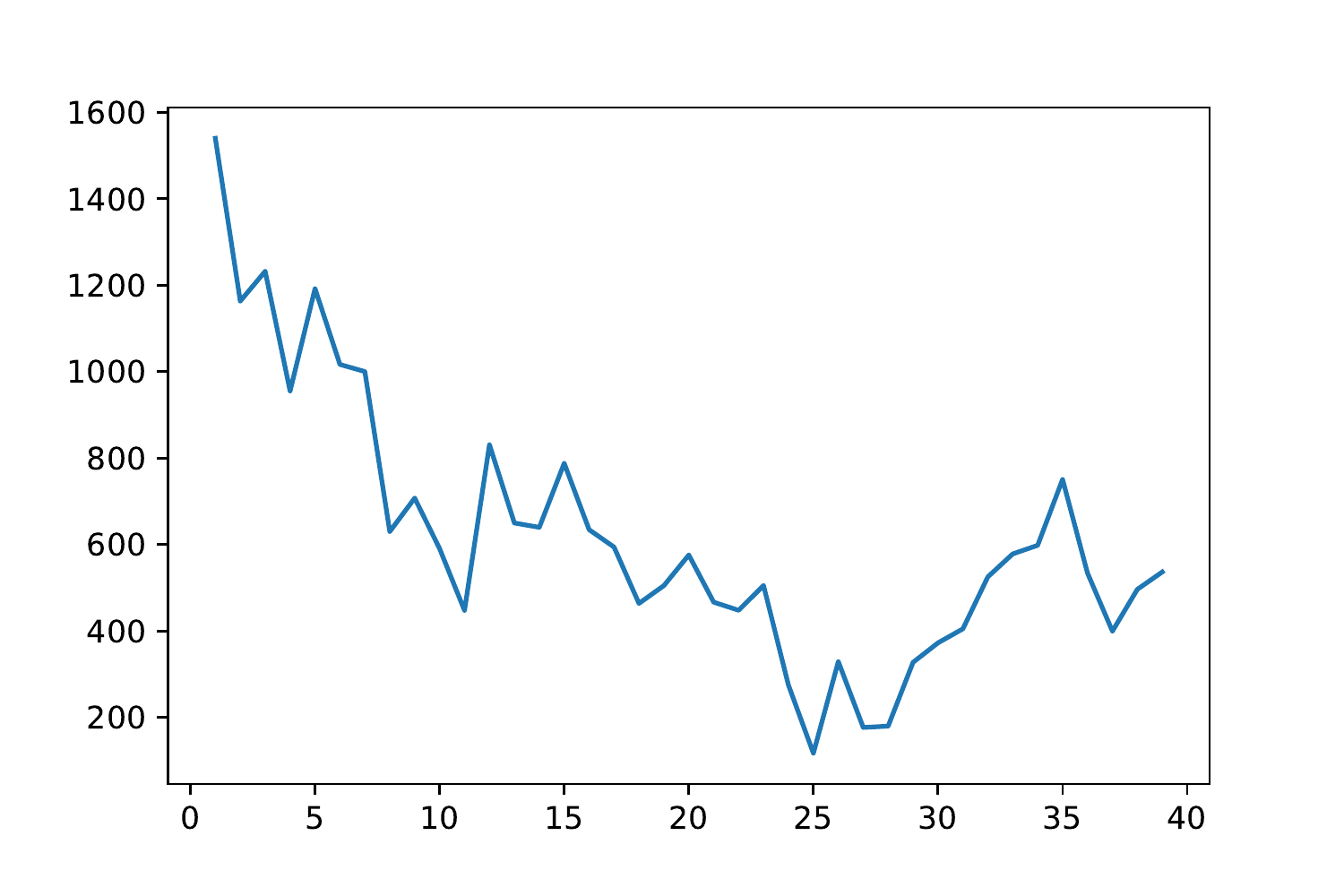}
  \label{subfigure: dayDFT_miss_size}
}
\subfloat[Data reuse counts]{%
  \includegraphics[width=0.24\linewidth, height=2.5cm]{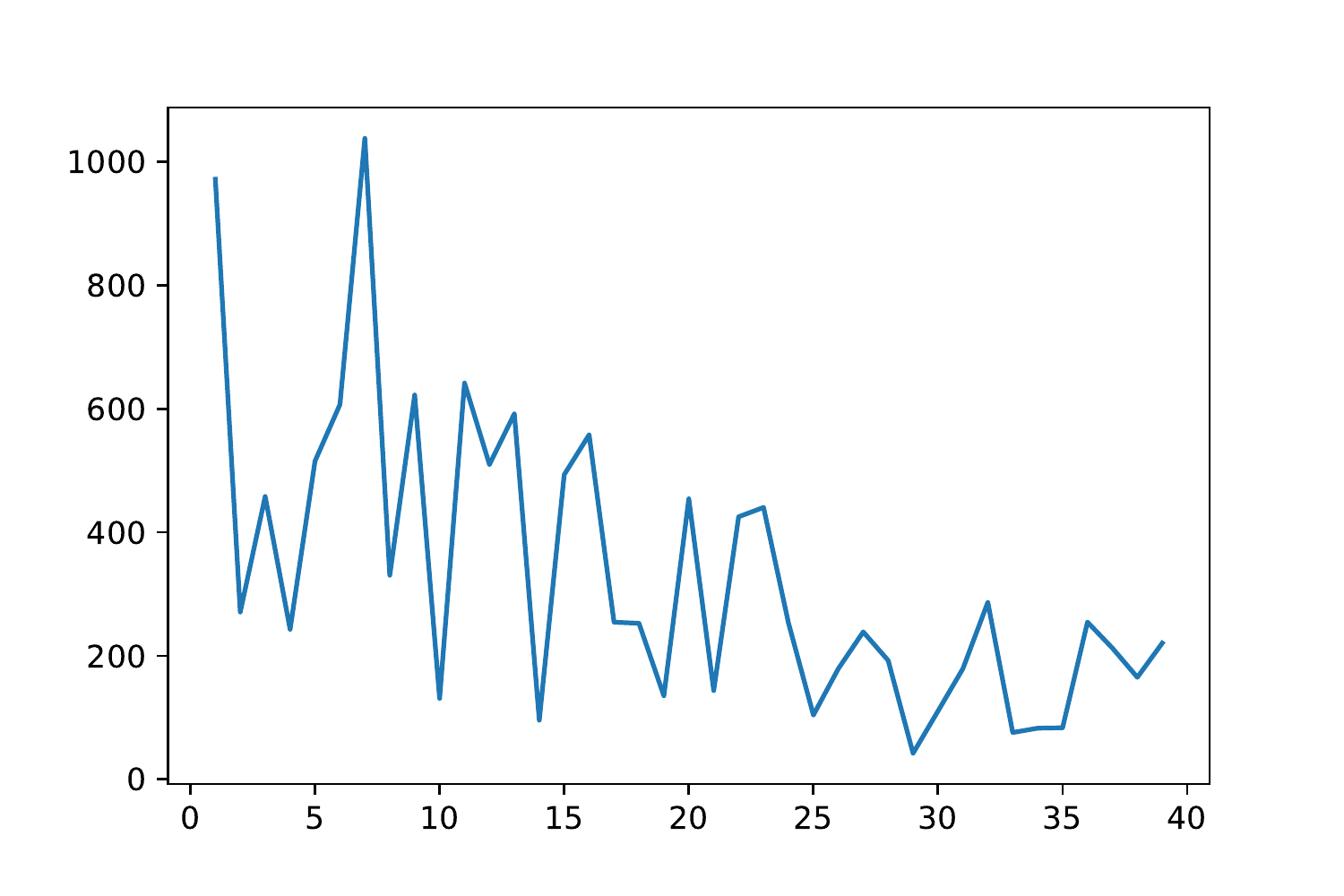}
  \label{subfigure: dayDFT_reuse_count}
}
\subfloat[Data reuse sizes]{%
  \includegraphics[width=0.24\linewidth, height=2.5cm]{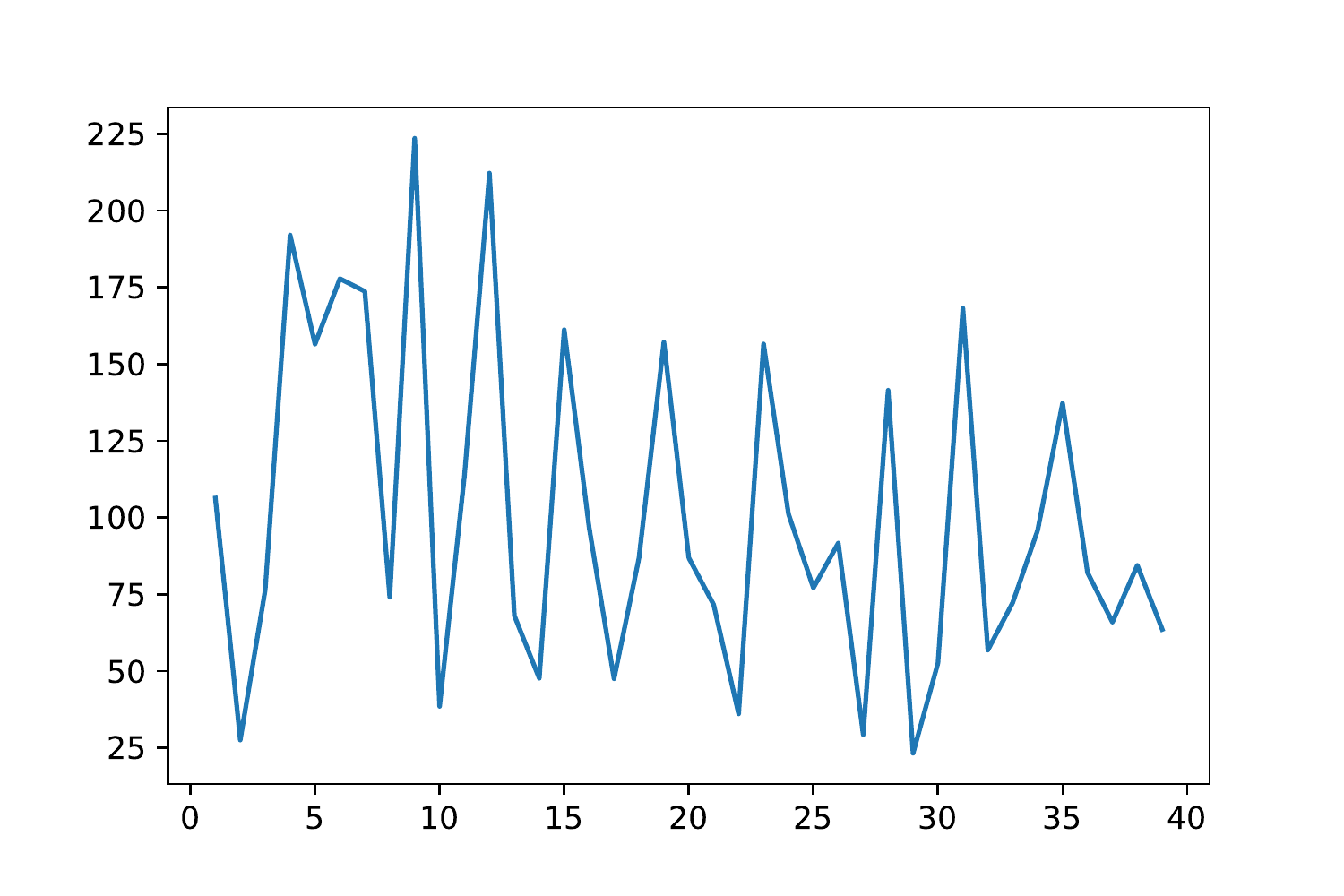}
  \label{subfigure: dayDFT_reuse_size}
} \newline
\caption{Periodogram of daily data.  All eight features show the same peaks at 31 days and 62 days.}
\vspace{-0.5cm}
\label{fig:dayDFT}
\end{figure*}

Figure \ref{fig:dayDFT} shows the periodogram of daily  data.
All columns show relatively strong, if not strongest, seasonal effects of 7 day period, confirming that there exists a weekly seasonal effect. %  The spikes happen at 31 and 62, indicating that there exist strong seasonal effects with the 31-day period and 62-day period. This shows the monthly and bi-monthly seasonality in the daily data.

\section{Conclusions}

In this paper, we studied the access trends of the Southern California Petabyte Scale Cache operated by teams of high-energy physicists in California.
Our analysis shows that the SoCal Repo was able to reduce the network traffic by 57\% for a large portion of the period of the study. However, some periods of study show access patterns of streaming data which is an inefficient way of using the caching system, and impacts the performance of the backbone network.
Through this study, we developed a number of machine learning models to further explore the predictability of the cache utilization statistics.
Because the regional storage cache could predictably reduce the network utilization, we anticipate that a more general caching mechanism could benefit many more scientific communities beyond the specific physics community studied.
%By studying the access patterns and potential for network traffic reduction by the caching system, we showed the predictability of the cache utilization, which would lead to the resource allocation planning for the network and storage. Our study showed that the network traffic volume was reduced by a factor of 2.35 before new cache nodes have been added, and by a factor of 1.30 overall during the study period. Long-term resource planning is a challenge, and the understanding characteristics of the access trends would provide the sustainability of the data access and insights of the challenge. Our LSTM model shows 0.884 of accuracy overall to predict cache utilization. As we collect more data, we plan to study the access trends for a longer period of time, and also enhance the network traffic prediction models.

The study also reveals a number of unexpected characteristics worth further investigation.
For example, the cache hit rates decrease significantly during the most recent months of the study, and a need for a larger dataset to train LSTM models. %the expected weekly periodicity is not visible in the periodograms.

\begin{acks}
This work was supported by the Office of Advanced Scientific Computing Research, Office of Science, of the U.S. Department of Energy under Contract No. DE-AC02-05CH11231, and also used resources of the National Energy Research Scientific Computing Center (NERSC). 
This work was also supported by the National Science Foundation through the grants OAC-2030508, OAC-1836650, MPS-1148698, PHY-1120138 and OAC-1541349.
\end{acks}

\bibliographystyle{ACM-Reference-Format}
\bibliography{main}

\end{document}